\documentclass[%
 reprint, 
 amsmath,amssymb,
 aps, 
 prd,
 floatfix,
 sort&compress,
 merge,
 longbibliography
]{revtex4-2}

\usepackage{graphicx}
\usepackage{dcolumn}
\usepackage{bm}

\usepackage[colorlinks,citecolor=blue,urlcolor=blue,linkcolor=blue]{hyperref}
\usepackage{color}
\usepackage{slashed}
\usepackage{multirow}

\usepackage{subfigure}
\usepackage{float}
\usepackage{amsmath}
\usepackage{bbm}

\newcounter{YJC}

\begin{document}

\title{Digit quantum simulation of a fermion field in an expanding universe} 

\author{Jia-Qi Gong}

\affiliation{Department of Physics, Liaoning Normal University, Dalian 116029, China}

\author{Ji-Chong Yang}
\email{yangjichong@lnnu.edu.cn}
\thanks{Corresponding author}

\affiliation{Department of Physics, Liaoning Normal University, Dalian 116029, China}
\affiliation{Center for Theoretical and Experimental High Energy Physics, Liaoning Normal University, Dalian 116029, China}

\date{\today}

\begin{abstract}
Quantum simulation is a rapidly evolving tool with great potential for research at the frontiers of physics, and is particularly suited to be used in computationally intensive lattice simulations, such as problems with non-equilibrium. 
In this work, a basic scenario, namely free fermions in an expanding universe, is considered and quantum simulations are used to perform the evolution and study the phenomena involved.
Using digital quantum simulations with the Jordan-Wigner transformation and Trotter decomposition, the evolutions of fermion number density, correlation functions, polarization, and chiral condensation are analyzed. 
A spread out phenomenon can be observed in the simulation, which is a consequence of momentum redshift. 
This work also demonstrates the simplicity and convenience of using quantum simulations when studying time-evolution problems.
\end{abstract}

\maketitle


\section{\label{sec:1}Introduction}

Quantum computing has emerged as a transformative tool in high-energy physics~(HEP) and cosmology, offering unprecedented capabilities to simulate complex quantum systems that are otherwise intractable for classical computers.
Although quantum computing is still in the era of Noisy Intermediate-Scale Quantum (NISQ) devices~\cite{Preskill:2018jim,Chen:2021num}, research in HEP based on quantum computing, is undergoing a rapid development phase~\cite{Fang:2024ple,Scott:2024txs,Roggero:2018hrn,Roggero:2019myu,Atas:2021ext,Lamm:2019uyc,Li:2021kcs,Perez-Salinas:2020nem,Jordan:2011ci,Mueller:2019qqj,Guan:2020bdl,Wu:2021xsj,Zhang:2023ykh,Zhang:2024ebl,Zhu:2024own,Fadol:2022umw,Wu:2020cye,Terashi:2020wfi,Yang:2024bqw}.
With the help of quantum superposition and entanglement, quantum computers can efficiently model phenomena such as particle interactions, quantum field dynamics, and spacetime curvature. 
This advancement holds the potential to revolutionize our understanding of fundamental physics, enabling precise simulations of quantum chromodynamics and other intricate quantum processes~\cite{Feynman:1981tf,Carena:2022kpg,Gustafson:2022xdt,Lamm:2024jnl,Carena:2024dzu,Li:2023vwx,Cui:2019sfz,Georgescu:2013oza,Echevarria:2020wct,Gong:2024iqu,Motta:2019yya,Czajka:2021yll}.
For instances, recent studies have demonstrated the application of quantum algorithms to simulate aspects of quantum field theories in curved spacetime and explore cosmological phenomena~\cite{Opanchuk:2013lgn,Bravo:2014twa,Yang:2019kbb,Bemani:2016rvi,Kinoshita:2024ahu,Maceda:2024rrd,Steinhauer:2015saa,Liu:2024wqj,Fulgado-Claudio:2022fut,Asaduzzaman:2023htk}.

Traditional computational methods face challenges known as the `sign problem' when simulating the real time evolution of quantum systems~\cite{Atas:2021ext,Lamm:2019uyc,Li:2021kcs,Carena:2022kpg}, particularly those involving fermion fields in dynamic spacetime backgrounds. 
The non-equilibrium nature of these systems, coupled with the need to account for quantum entanglement and particle creation, renders classical simulations computationally intensive. 
Quantum computers, with their ability to naturally represent and manipulate quantum states, offer a promising avenue to overcome these limitations. 

One example of a time evolution system is the fermion fields in an expanding universe.
Such scenario is crucial for comprehending the early universe's evolution, including the processes of inflation and reheating~\cite{Kumar:2018rrl} and can also affect quantum entanglement of particles~\cite{Li:2023gtf,Wu:2022glj,Wu:2022lmc,Wu:2024yop}. 
Fermions, as the building blocks of matter, play a pivotal role in the universe's structure formation and the synthesis of elements. 
While significant progress has been made in understanding these processes and analog experiments such as cold-atom systems~\cite{Steinhauer:2021fhb,Viermann:2022wgw,Hu:2018psq}, a relatively paucity of research has been conducted on this subject employing digital quantum simulations on universal quantum computers which operate via gate-based, programmable architectures that can be reconfigured to simulate a wide variety of quantum systems, and are able to provide advantages with the future development of algorithms such as the error corrections~\cite{Temme:2016vkz,Kandala:2018kwe,Ni:2022cdv,GoogleQuantumAIandCollaborators:2024efv,Zhao:2021chc} and circuit optimizations~\cite{Keever:2022hfy,Zou:2021pvl}. 
As a proof of concept, this work investigates the free fermion field in the $1+1$ dimensional expanding universe using a digital quantum simulation.

The remainder of the paper is organized as follows.
In Section~\ref{sec:2}, the model to be simulated is presented.
The simulation and numerical results are shown in section~\ref{sec:3}.
Section~\ref{sec:4} is a summary of the conclusions.

\section{\label{sec:2}The model}

The metric in 2-dimensional Friedmann-Lema\'{\i}tre-Robertson-Walker spacetime~\cite{Friedman:1922kd,Lemaitre:1931zzb,Robertson:1935zz,Robertson:1935jpx,Robertson:1936zz,Walker:1937qxv} is,
\begin{equation}
d s^{2}=d t^{2}-g^{2}(t) d x^{2}
\label{eq.2.1}
\end{equation}
where $g\left ( t \right )$ is the scale factor describing the expansion of the universe. $t$ is the cosmic time coordinate and $x$ is the spatial coordinate.
The corresponding Hamiltonian of a free fermion is~(details are shown in Appendix~\ref{sec:ap1}),
\begin{equation}
H=a \sum_{x} \bar{\psi}\left(-i \gamma^{1} \partial_{x}+\frac{1}{2}\frac{g'(t)}{g(t)}\gamma^{0}+g(t) m\right) \psi, 
\label{eq.2.2}
\end{equation}
where $a$ is lattice spacing, and $\gamma$ matrices in $1+1$ dimension are $\gamma _0=\sigma ^z$, $\gamma _1=-i\sigma ^y$ where $\sigma ^i$ are Pauli matrices. 
Denoting $X=2x$ as the even sites, the Dirac fermion field $\psi (X)$ can be written with the staggered fermion field $\chi(x)$ as~\cite{Czajka:2021yll,Kogut:1974ag,Kluberg-Stern:1983lmr,Morel:1984di}, 
\begin{equation}
\psi(X)=\frac{1}{\sqrt{a}}\left(\begin{array}{c}\chi (X) \\ \chi(X+1) \end{array}\right),
\label{eq.2.4}
\end{equation}
where the factor $1/\sqrt{a}$ is added to make $\chi(x)$ dimensionless.
In the case of the derivation of $\psi$, we use forward derivation and backward derivation for different components of the spinor,
\begin{equation}
\begin{split}
&\partial _X \psi(X) = \frac{1}{a\sqrt{a}} \begin{pmatrix} \chi (X+2) -\chi (X) \\ \chi (X+1) - \chi (X-1) \end{pmatrix}.
\end{split}
\label{eq.2.5}
\end{equation}
In the Jordan-Wigner representation~\cite{JW},
\begin{equation}
\chi(x)=\frac{\sigma^x(x)-{\rm i} \sigma^y(x)}{2} \prod_{j=0}^{x-1}\left(-{\rm i} \sigma^z(j)\right),
\label{eq.2.6}
\end{equation}
where $\sigma ^i(x)$ are Pauli matrices sitting on sites but not in the spinor space.

To transform the Hamiltonian into the Jordan-Wigner representation, one needs the following results of bilinear terms obtained when $N$ is even and by using Eqs~(\ref{eq.2.4})-(\ref{eq.2.6}), and with periodic boundary condition,
\begin{equation}
\begin{split}
 & a \sum_X \bar{\psi}(X) i \gamma_1 \partial_X \psi(X)\\
 &=\frac{1}{2 a} \sum_{x=0}^{N-2}\left(\sigma^x(x) \sigma^x(x+1)+\sigma^y(x) \sigma^y(x+1)\right) \\ 
 & +\frac{(-1)^{\frac{N}{2}}}{2 a} \prod_{j=1}^{N-2} \sigma^z(j)\\
 &\times \left(\sigma^x(0) \sigma^x(N-1)+\sigma^y(0) \sigma^y(N-1)\right),\\
&a \sum_X \bar{\psi}(X) \gamma_0 \psi(X)=\sum_{x=0} \frac{1+\sigma^z(x)}{2},\\
&a \sum_X \bar{\psi}(X) \psi(X)=\sum_{x=0}(-1)^x \frac{1+\sigma^z(x)}{2}.
\end{split}
\label{eq.2.7}
\end{equation}

In the following, we consider a de Sitter space~\cite{DeSitter} with $g(t)=e^{ht}$, where $h$ is the Hubble constant, the Hamiltonian is then,
\begin{equation}
\begin{split}
 & aH=-\frac{1}{2} \sum_{x=0}^{N-2}\left(\sigma^x(x) \sigma^x(x+1)+\sigma^y(x) \sigma^y(x+1)\right) \\ 
 & -\frac{(-1)^{\frac{N}{2}}}{2} \prod_{j=1}^{N-2} \sigma^z(j)\\
 &\times \left(\sigma^x(0) \sigma^x(N-1)+\sigma^y(0) \sigma^y(N-1)\right)\\
 &+\frac{ah}{4}\sum_{x=0}^{N-1}\sigma^z(x)+\frac{ame^{ht}}{2}\sum_{x=0}^{N-1}(-1)^x\sigma^z(x).\\
\end{split}
\label{eq.2.8}
\end{equation}

\section{\label{sec:3}Simulation of the system}

\subsection{\label{sec:3.1}set up of the simulation}

The fermion number in an FLRW universe can be defined as,
\begin{equation}
\begin{split}
&n=\int dX \sqrt{-\det(g_{\mu\nu})}\psi^{\dagger}\psi ,
\end{split}
\label{eq.3.1}
\end{equation}
which is discretized as,
\begin{equation}
\begin{split}
&\hat{n}(x)=\chi^{\dagger}(x) \chi (x) =\frac{1+\sigma ^z\left (x\right )}{2},\;\;\hat{n}(t)=e^{ht}\sum _x n(x),
\end{split}
\label{eq.3.2}
\end{equation}
It can be verified that $[\sum_x\sigma^z(x),H]=0$, therefore $\sum_x\sigma^z(x)$ is conserved, and $\langle \hat{n}(t)\rangle = e^{ht} n_0$, where $n_0$ is the fermion number at $t=0$.
Note that the volume also scales the same, rendering an unchanged particle density.
Meanwhile, the spatial distribution of $n$ is not conserved, which is the quantity to be studied numerically.

If the initial states are chosen from the set $|k\rangle$, it can be shown that $|0\rangle$ and $|2^N-1\rangle$ are eigen-states of Hamiltonian with $H(t)|0\rangle = \left(Nh/4\right)|0\rangle$, and $H(t)|2^N-1\rangle = \left(-Nh/4\right)|2^N-1\rangle$. 
As a consequence, $e^{iH(t)t}|0\rangle$ and $e^{iH(t)t}|2^N-1\rangle$ are different from $|0\rangle$ and $|2^N-1\rangle$ by global phases only, and therefore all observables remain constants.
For the other $|k\rangle$, this is not the case.
By using the definition of the fermion number, it can be recognized that, $|0\rangle$ and $|2^N-1\rangle$ correspond to full filling and zero filling states, respectively.

In this work, $|1\rangle$ and $|2^N-2\rangle$ are chosen as the initial states. 
$|1\rangle$ can be recognized as a state with fermions occupying every lattice site except for the first site, while $|2^N-2\rangle$ can be recognized as a state with a fermion occupying the first site.

Apart from the fermion number, other observables are also studied, including the density correlation, which is defined as $\hat{C}(x-y)= \hat{n}(x)\hat{n}(y)$, where $\hat{n}$ is the fermion number defined in Eq.~(\ref{eq.3.2}), taking $x=0$ and $y=1$, 
\begin{equation}
\begin{split}
&\hat{C}=\frac{1+\sigma^z(0)\sigma^z(1)+\sigma^z(0)+\sigma^z(1)}{4}.\\
\end{split}
\label{eq.3.3}
\end{equation}
The electric polarization of the $\chi$ field is also of interest which can be defined as $p=e\int dx \sqrt{-\det(g_{\mu\nu})} x\chi^{\dagger}(x)\chi(x)$, where $e$ is the electric charge, and can be discretized as,
\begin{equation}
\begin{split}
&\frac{\hat{p}(t)}{e}= e^{ht}\sum_{x=0}x\chi ^{\dagger}(x)\chi (x) = e^{ht}\sum_{x=0}x \frac{1+\sigma^z(x)}{2}.
\end{split}
\label{eq.3.4}
\end{equation}
Another important quantity is the chiral condensation which serves as the order parameter of a chiral symmetry breaking phase transition, and is defined as,
\begin{equation}
\begin{split}
&c(t)=\int dx \sqrt{-\det(g_{\mu\nu})}\bar{\psi}\psi ,\\
&\hat{c}(t)= e^{ht}\sum_{x=0}(-1)^x \frac{1+\sigma^z(x)}{2}.
\end{split}
\label{eq.3.5}
\end{equation}

The simulation is implemented using the \verb"Qiskit" toolkit~\cite{qiskit2024}, and carried out with $N=8$~(the Hamiltonian when $N=8$ is shown in Appendix~\ref{sec:ap2}).
A Trotter decomposition~\cite{Lloyd:1996aai} is applied with $K=40$ steps, i.e. for $H(t)=\sum _j a_j(t) \hat{\sigma}_j$ where $\hat{\sigma}_j$ are tensor products of Pauli matrices, $a_j(t)$ are real coefficients depending on time, $e^{iH(t)t}$ is approximated as,
\begin{equation}
\begin{split}
&e^{iHt}\approx \prod _j e^{i a_j((K-1)\Delta t)\Delta t \hat{\sigma}_j} \times \prod _j e^{i a_j((K-2)\Delta t)\Delta t \hat{\sigma}_j} \times\\
&\ldots \times \prod _j e^{i a_j(0)\Delta t \hat{\sigma}_j},
\end{split}
\label{eq.3.6}
\end{equation}
where $\Delta t=t/K$.
In the simulation, $h=0.1 a^{-1}$, $am=0, 1$ and $t=a$~(therefore for each Trotter step $\Delta t=0.1a$), where $a$ is the lattice spacing.
The observables of interest in this study are measured for $r=50000$ times.
With $r$ repetitions, the statistical errors of measurements can also be obtained,
\begin{equation}
\begin{split}
&\varepsilon_{n(x)}=\sqrt{\frac{\langle \hat{n}(x)\rangle(1-\langle \hat{n}(x)\rangle)}{r}},\\
&\varepsilon_{C(x,y)}=\sqrt{\frac{\langle \hat{C}\rangle (1-\langle \hat{C} \rangle)}{r}},\\
&\varepsilon_{p(t)/e}=\frac{e^{ht}}{2\sqrt{r}}\sqrt{\sum _{x=1}^{N-1}x^2(1-\langle\sigma ^z(x)\rangle ^2)},\\
&\varepsilon_{c(t)}=\frac{e^{ht}}{2\sqrt{r}}\sqrt{\sum _{x=0}^{N-1}(1-\langle\sigma ^z(x)\rangle ^2)}.\\
\end{split}
\label{eq.3.7}
\end{equation}
In the presentation of numerical results, the statistical errors are also included.

\subsection{\label{sec:3.2}exact diagonalization}

Since $N=8$, $H$ is a $256\times 256$ matrix, the exact diagonalization can be calculated.
The results calculated by using exact diagonalization are also present as a comparison to study the deviation due to Trotter decomposition.
For an initial state $|\phi\rangle$, the result of $|\phi(t)\rangle$ is obtained as,
\begin{equation}
\begin{split}
&| \phi(t=K\Delta t) \rangle = \prod _{k=K-1,K-2,\ldots,0} e^{iH(k\Delta t)\Delta t}|\phi \rangle,
\end{split}
\label{eq.3.8}
\end{equation}
where the definitions of $K$ and $\Delta t$ are as same as in Eq.~(\ref{eq.3.6}), but the values are different.

\begin{figure}[htbp]
\includegraphics[width=0.48\hsize]{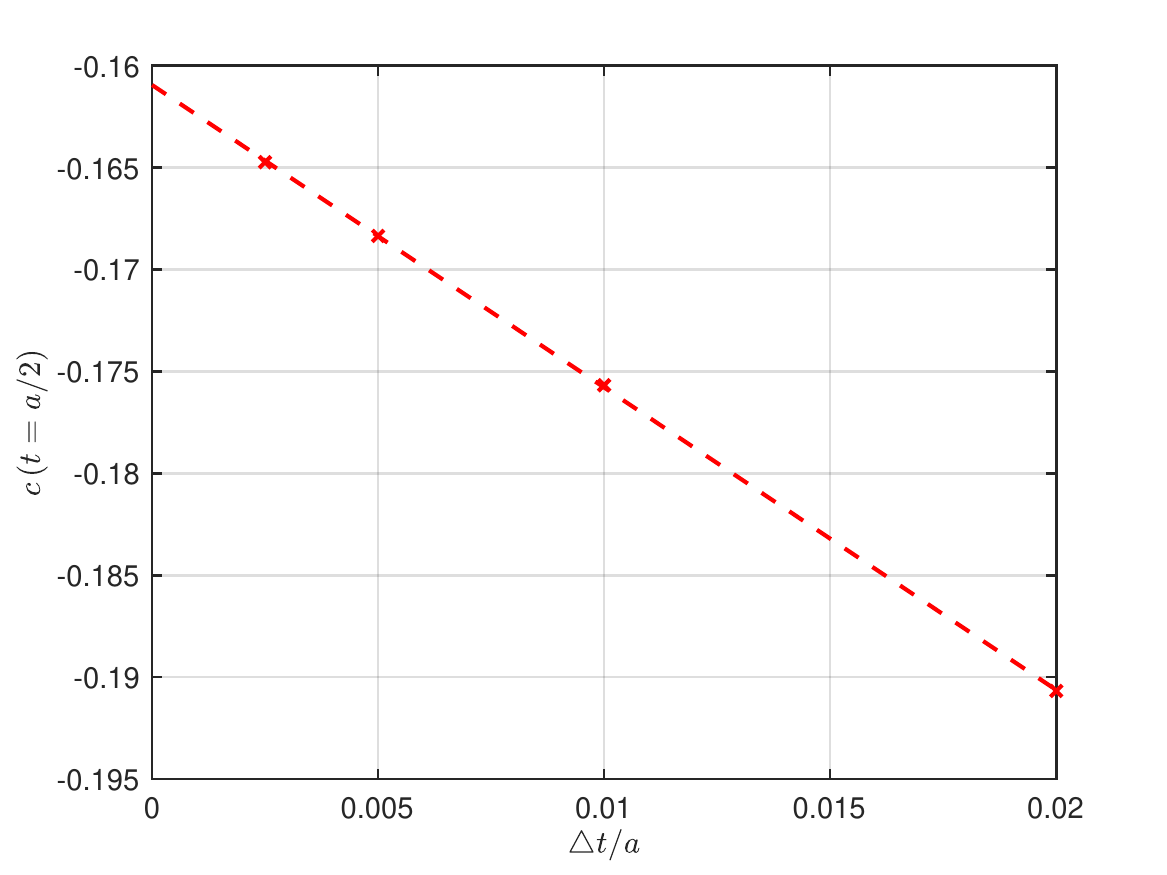}
\includegraphics[width=0.48\hsize]{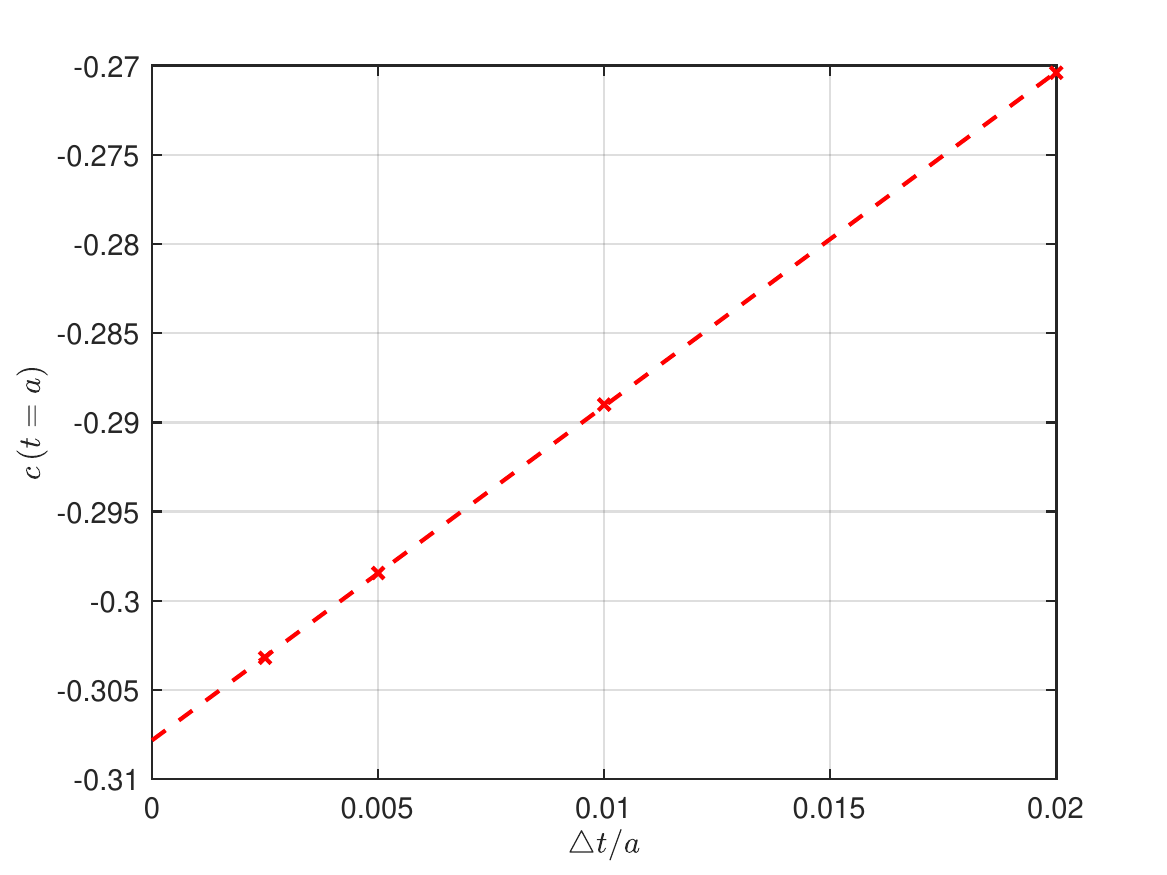}\\
\caption{\label{fig:extrapolationexamples}Results of $c(t=a/2)$~(left panel) and $c(t=a)$~(right panel) by using exact diagonalization for different $\Delta t$ as well as the linear extrapolations.
It can be found that, the linear extrapolation is able to obtain the results at $\Delta t\to 0$.}
\end{figure}
\begin{figure}[htbp]
\includegraphics[width=0.7\hsize]{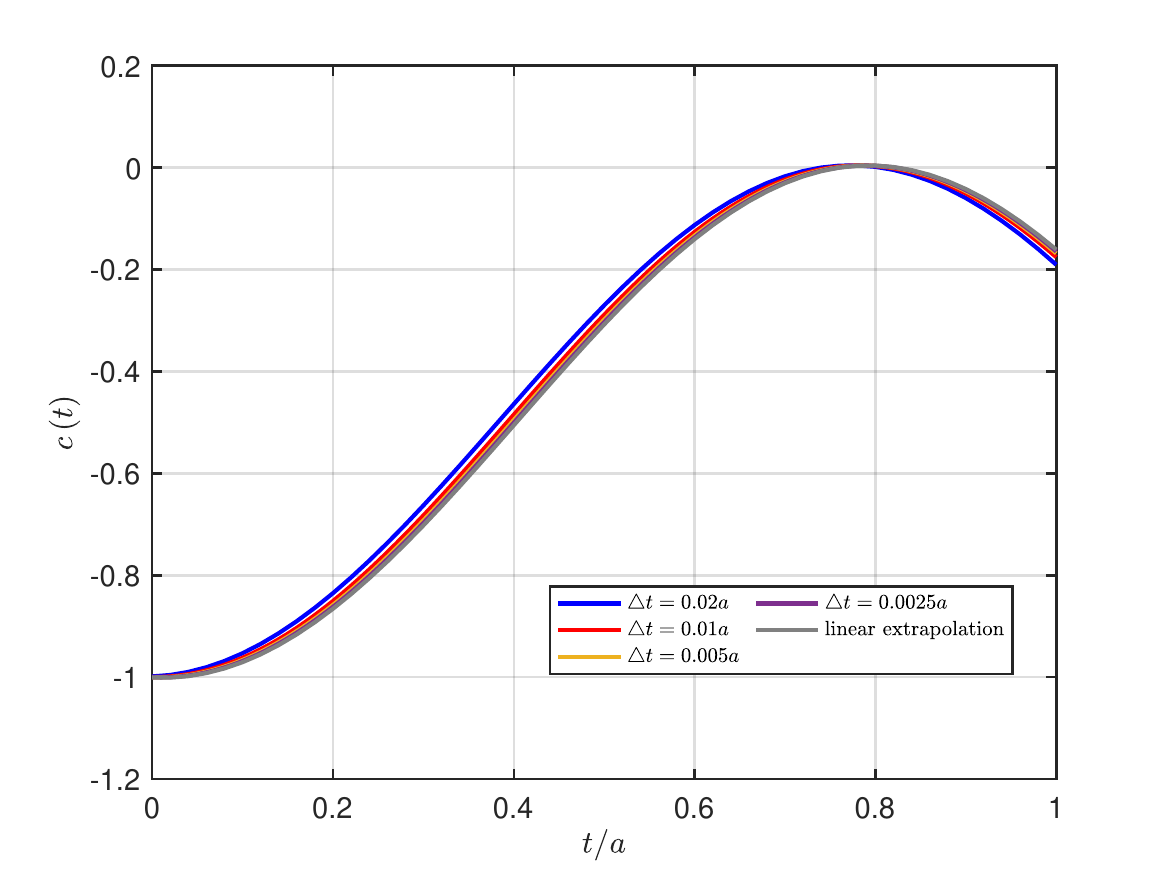}
\caption{\label{fig:extrapolationresult}$c(t)$ as functions of $t$ for different $\Delta t$ compared with $c(t)$ obtained by using linear extrapolation.}
\end{figure}
Taking the case of chiral condensation with the initial state $|1\rangle$ and with $am=1$ as an example, denoting $c(t)=\langle \hat{c}(t)\rangle _t$ where the subscript $t$ means measured using $\phi(t)$ and $\hat{c}(t)$ is defined in Eq.~(\ref{eq.3.5}), the result with $c(t=a/2)$ and $c(t=a)$ for different $\Delta t$ are studied.
The result with $\Delta t=0.0025$, $0.005$, $0.01$ and $0.02$~($K=400$, $200$, $100$ and $50$, respectively) are shown in Fig.~\ref{fig:extrapolationexamples}, with the linear extrapolation.
It can be seen that, the extrapolation works fine to obtain the result with $\Delta t\to 0$.
In this work, results with $\Delta t=0.0025$, $0.005$, $0.01$ and $0.02$ are calculated. 
Finally a linear extrapolation to $\Delta t\to 0$ is performed to obtain the result of exact diagonalization.
The example of chiral condensation as functions of $t$ for different $\Delta t$ and $c(t)$ obtained by the linear extrapolation are compared in Fig.~\ref{fig:extrapolationresult}.

\subsection{\label{sec:3.3}numerical results}

\begin{figure}[htbp]
\includegraphics[width=0.48\hsize]{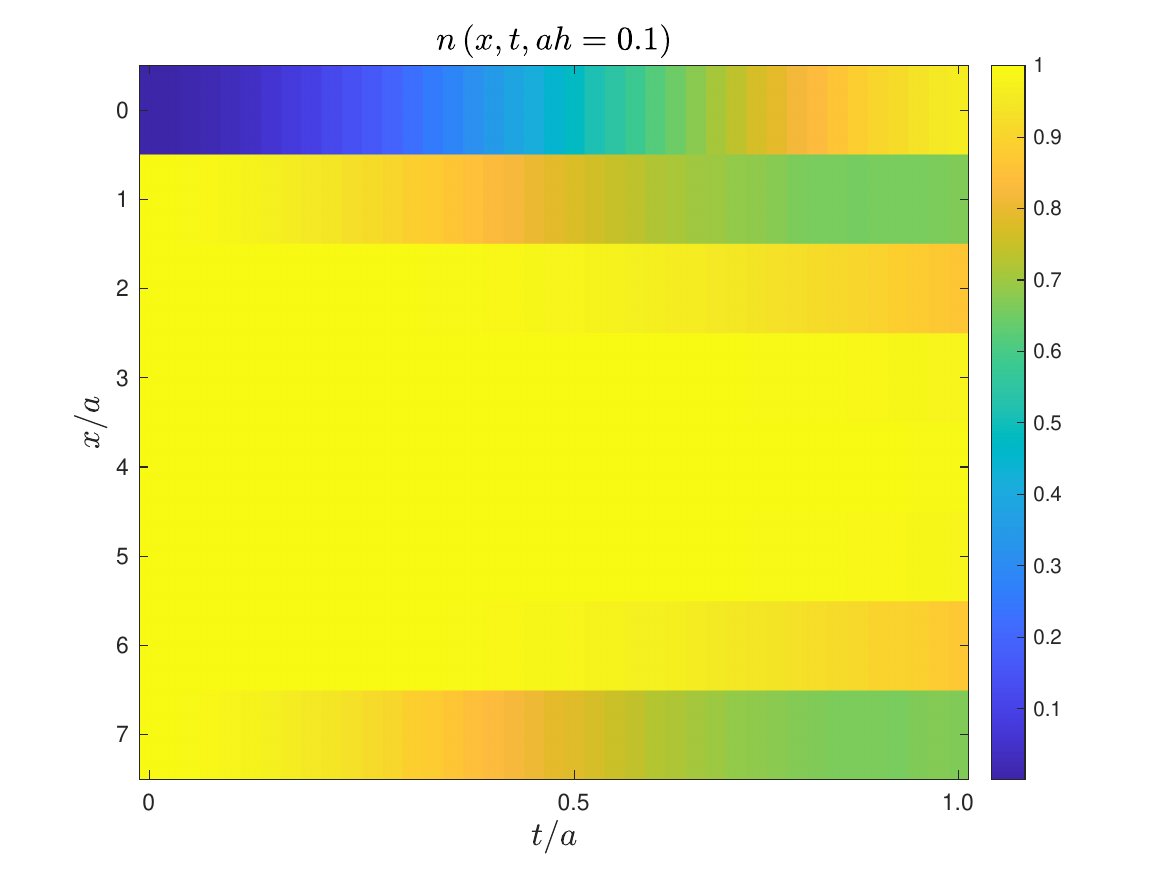}
\includegraphics[width=0.48\hsize]{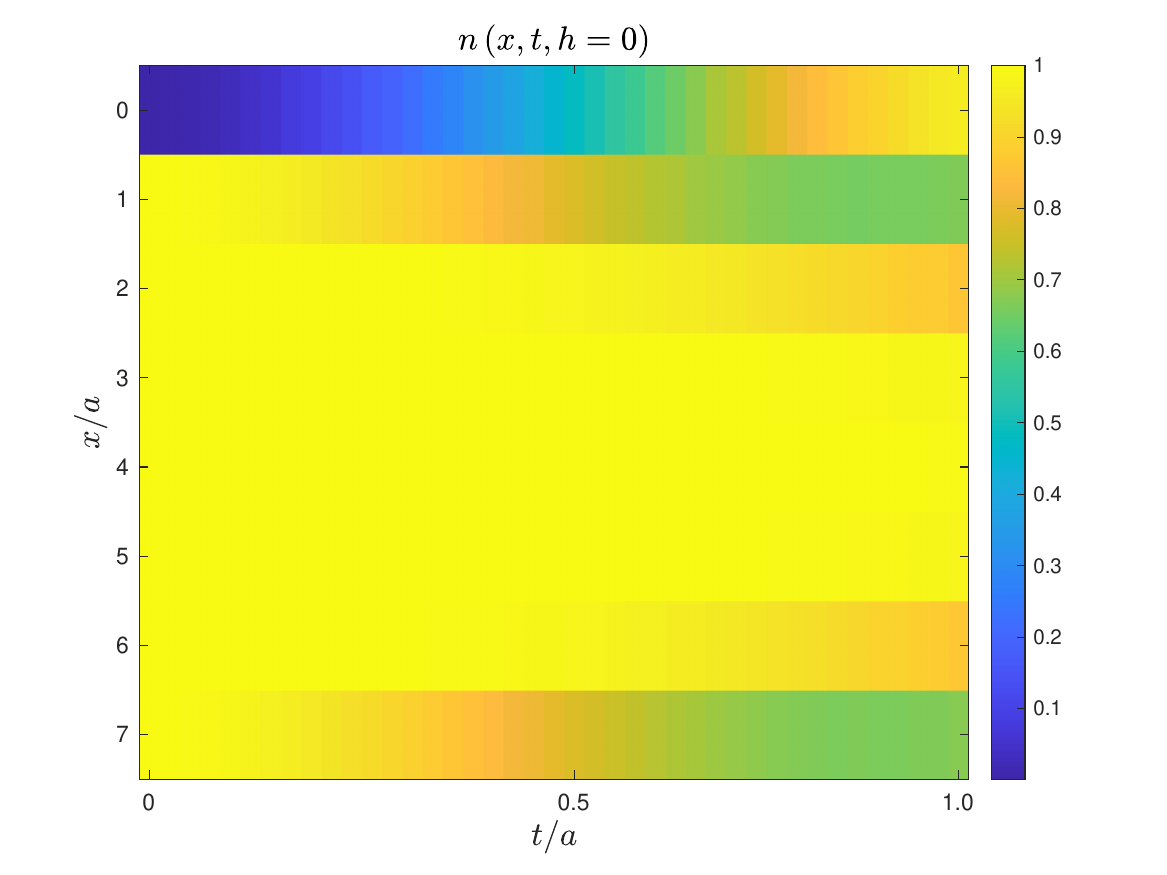}\\
\includegraphics[width=0.48\hsize]{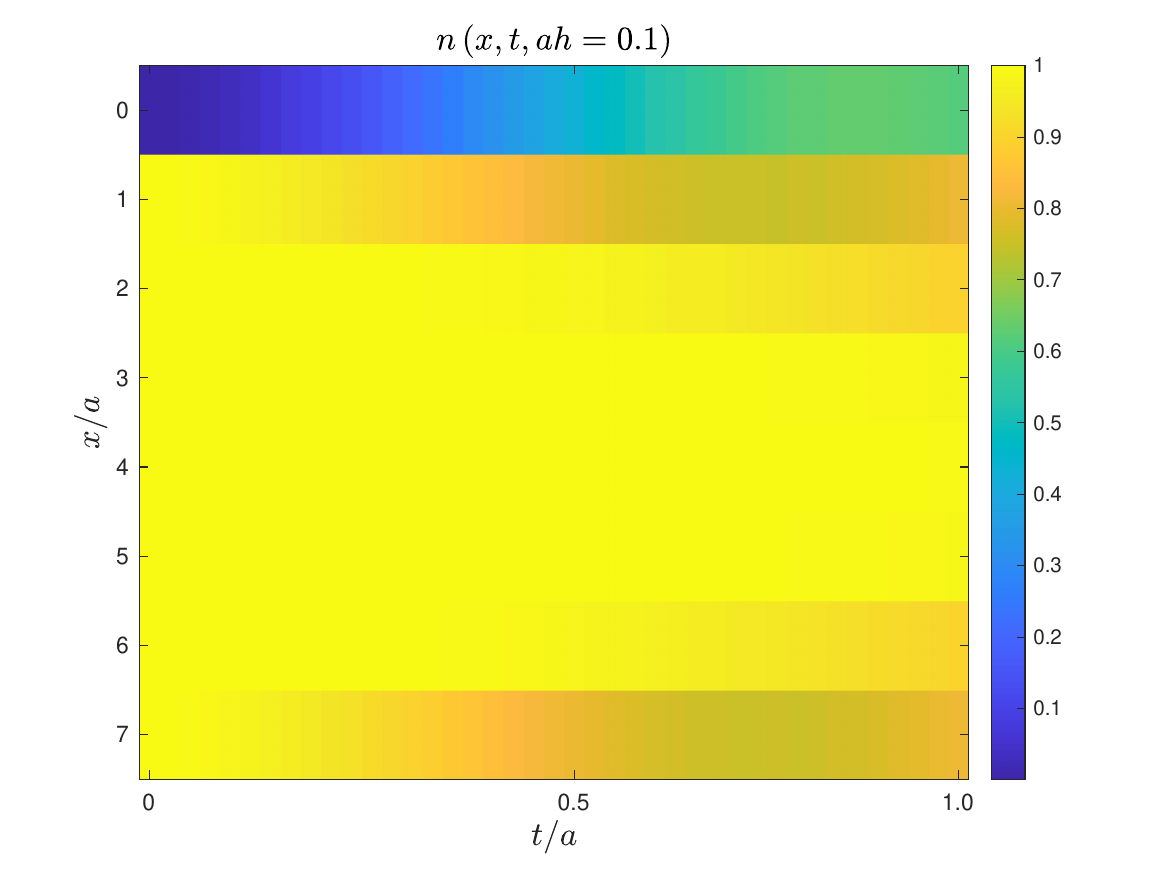}
\includegraphics[width=0.48\hsize]{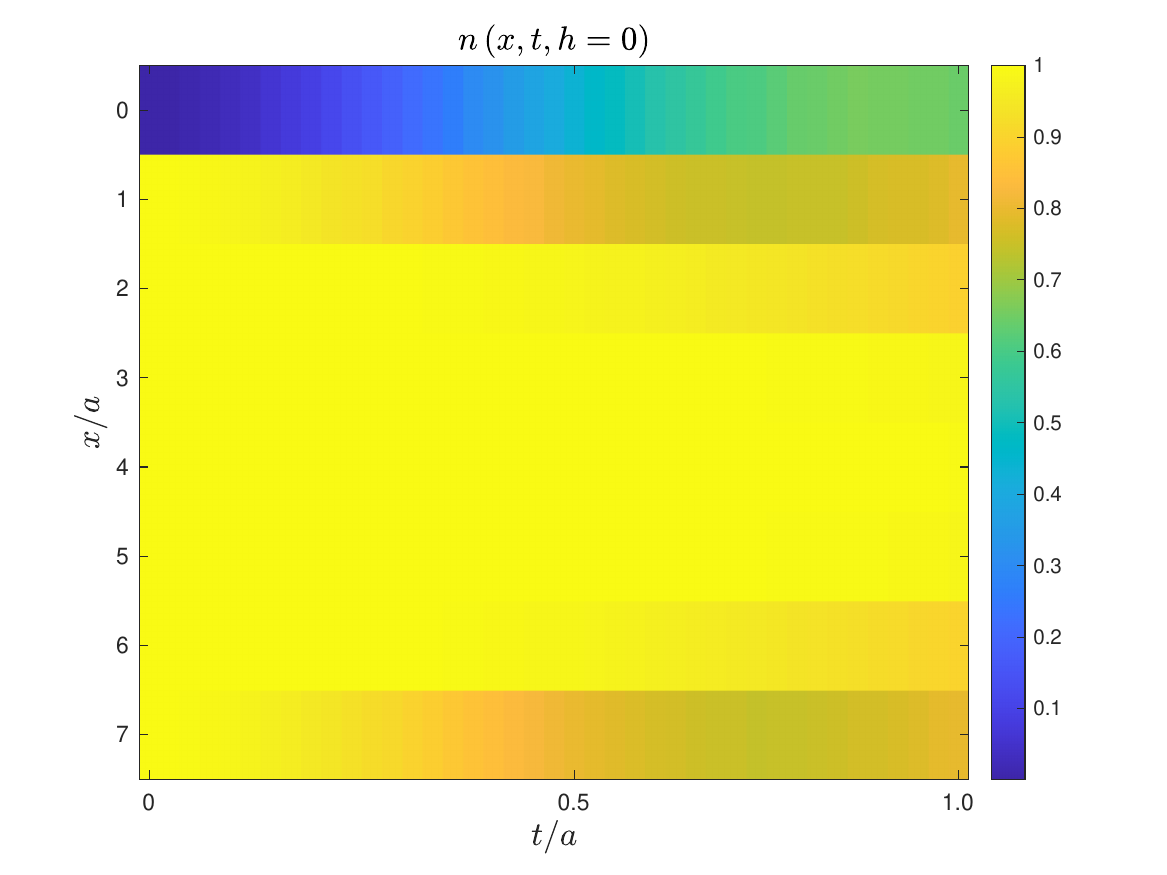}\\
\caption{\label{fig:fermionnumberstate1}Starting from initial state $|1\rangle$, fermion number distribution changes with time evolution.
The upper panels correspond to $m=0$, and the lower panels correspond to $am=1$, respectively.
The left panels correspond to $ah=0.1$, and the right panels correspond to $h=0$, respectively.}
\end{figure}

\begin{figure}[htbp]
\includegraphics[width=0.48\hsize]{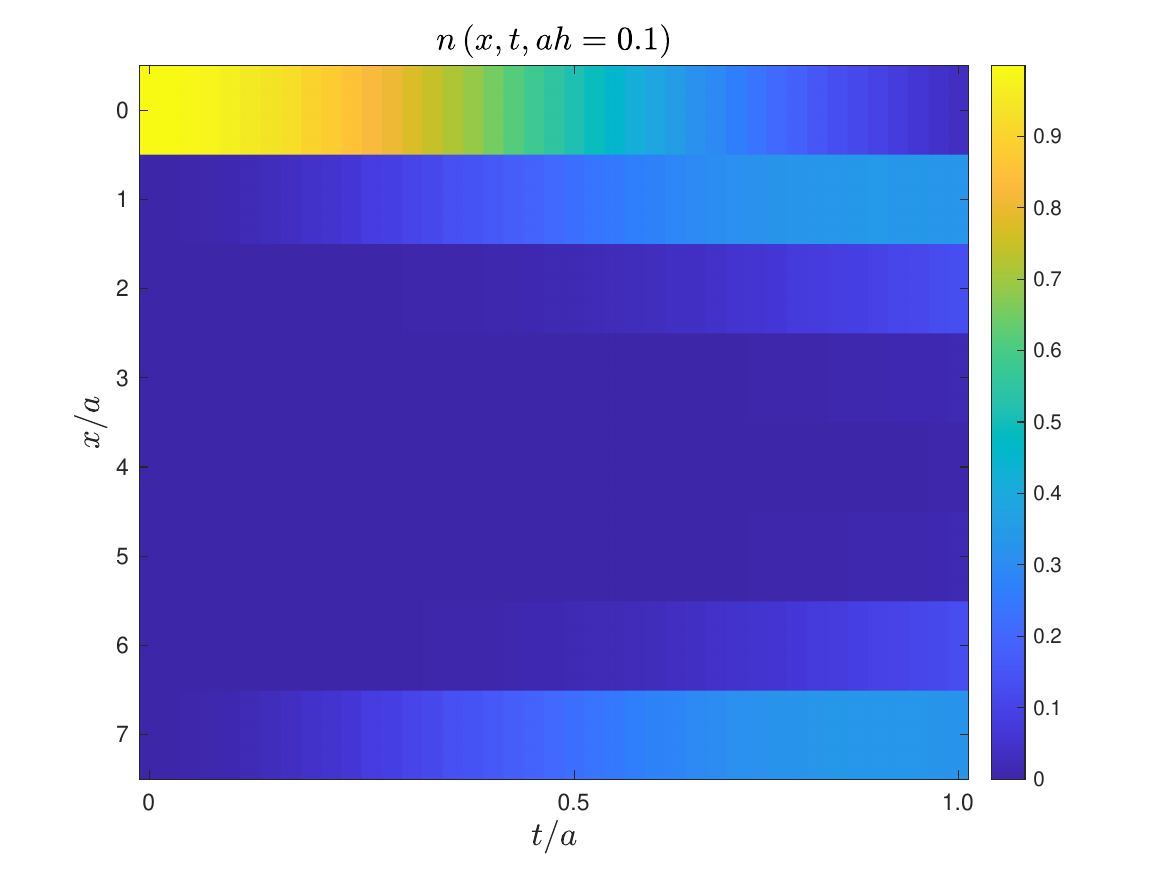}
\includegraphics[width=0.48\hsize]{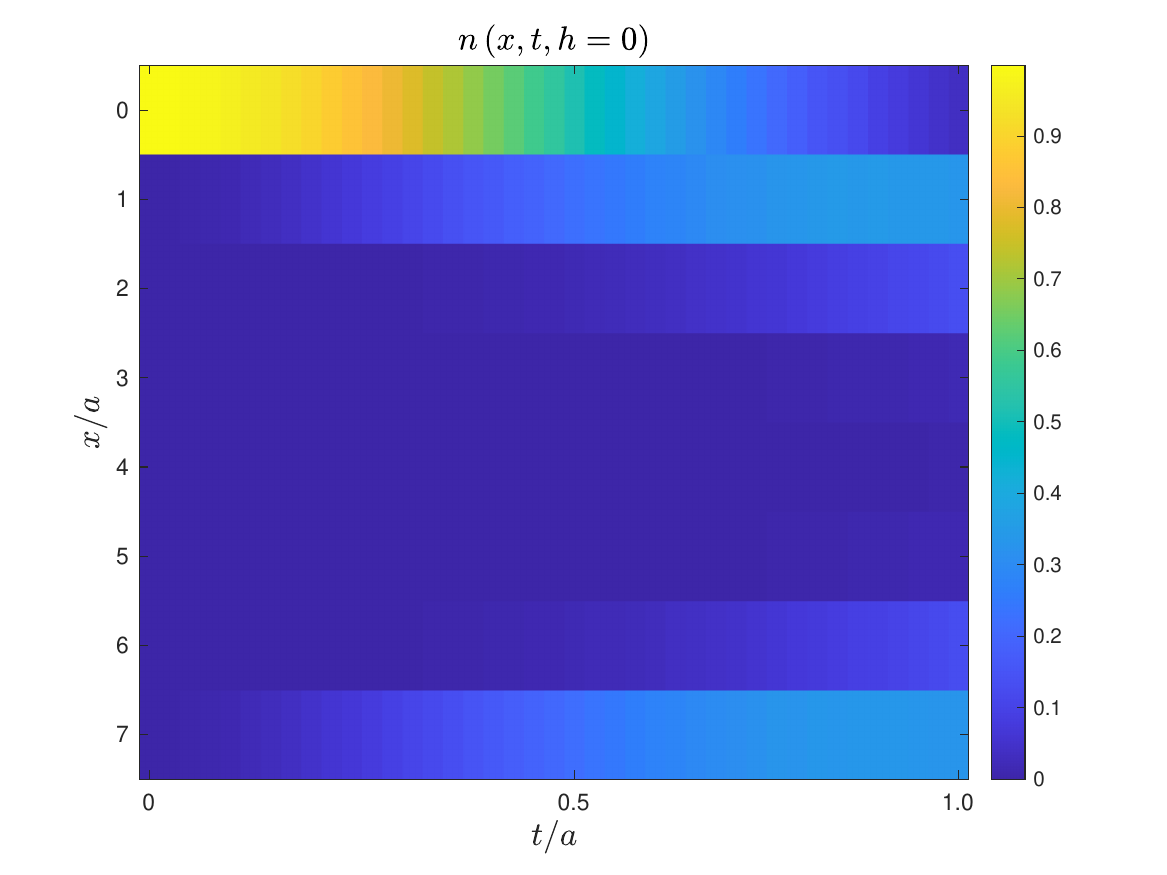}\\
\includegraphics[width=0.48\hsize]{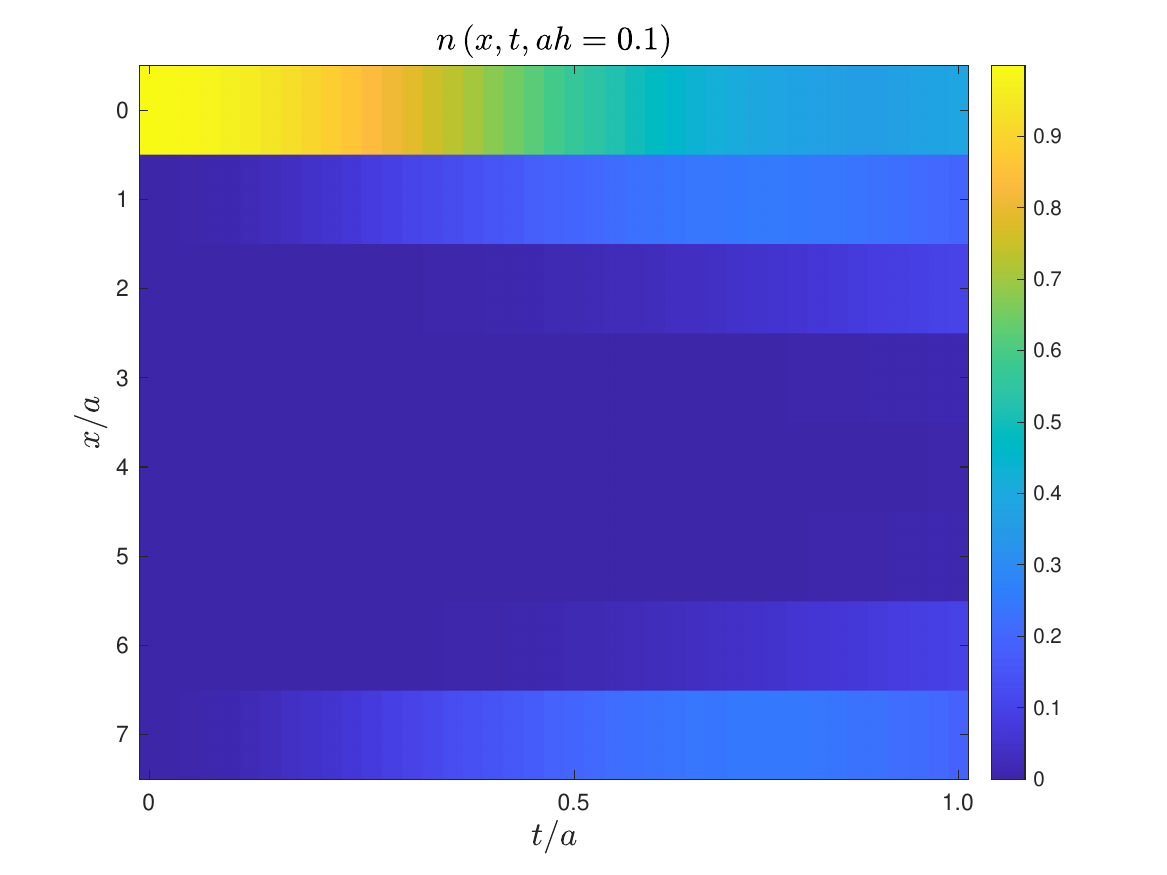}
\includegraphics[width=0.48\hsize]{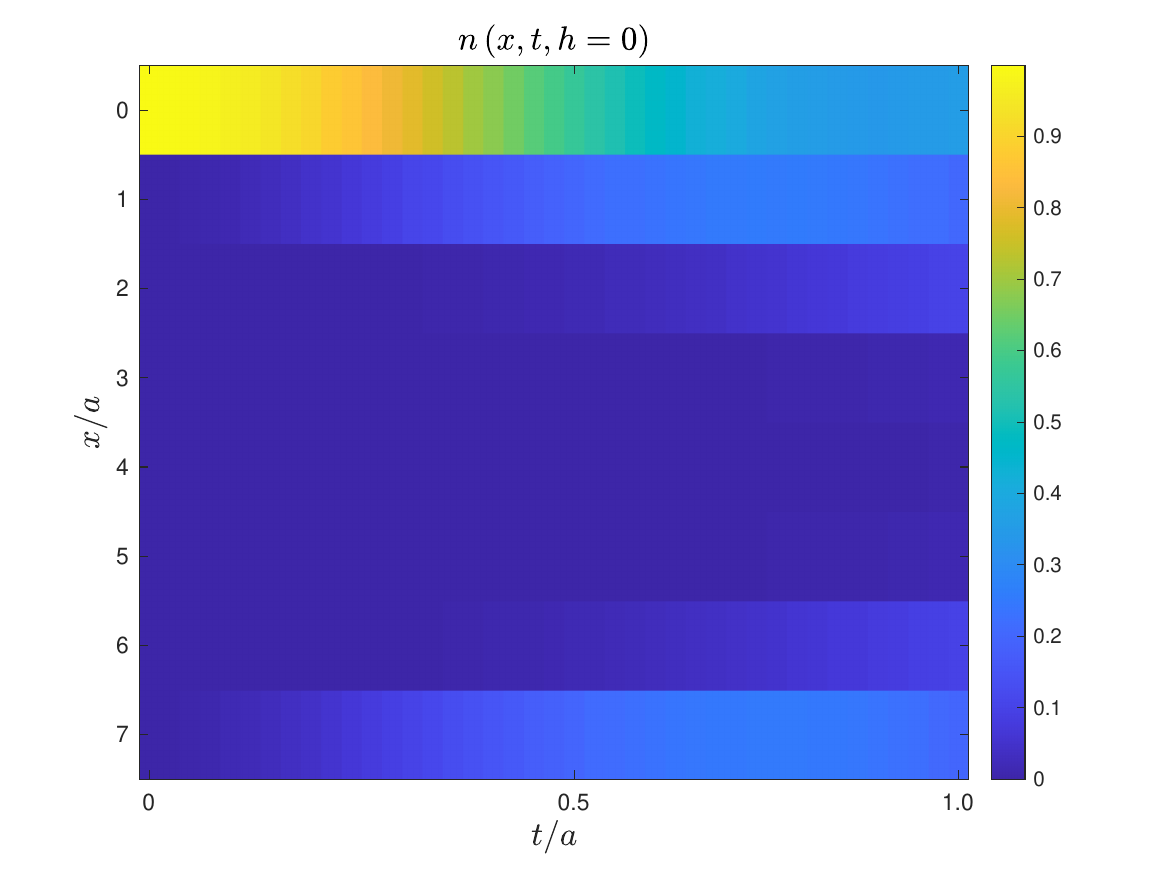}\\
\caption{\label{fig:fermionnumberstate2}Same as Fig.~\ref{fig:fermionnumberstate1} but for initial state $|254\rangle$.}
\end{figure}
We first focus on the fermion number. 
Denoting $n(x,t,h)=\langle \hat{n}(x)\rangle _{h,t}$, where the subscript `$h,t$' means measured using $|\phi(t)\rangle$ evaluated with the Hamiltonian corresponding Hubble constant $h$, and $\hat{n}(x)$ is defined in Eq.~(\ref{eq.3.2}), for the two initial states $|1\rangle$ and $|254\rangle$~(which is $2^N-2$ with $N=8$), the cases with and without expansion are shown in Figs.~\ref{fig:fermionnumberstate1} and \ref{fig:fermionnumberstate2}, respectively. 
It can be seen that wave packet spreading occurs regardless of whether there is expansion.
In fact, the difference between the cases of $h\neq 0$ and $h=0$ is very small.
To study the effect of an expanding universe, we subsequently focus primarily on the differences between the cases with expansion and the cases without expansion.

\begin{figure}[htbp]
\includegraphics[width=0.48\hsize]{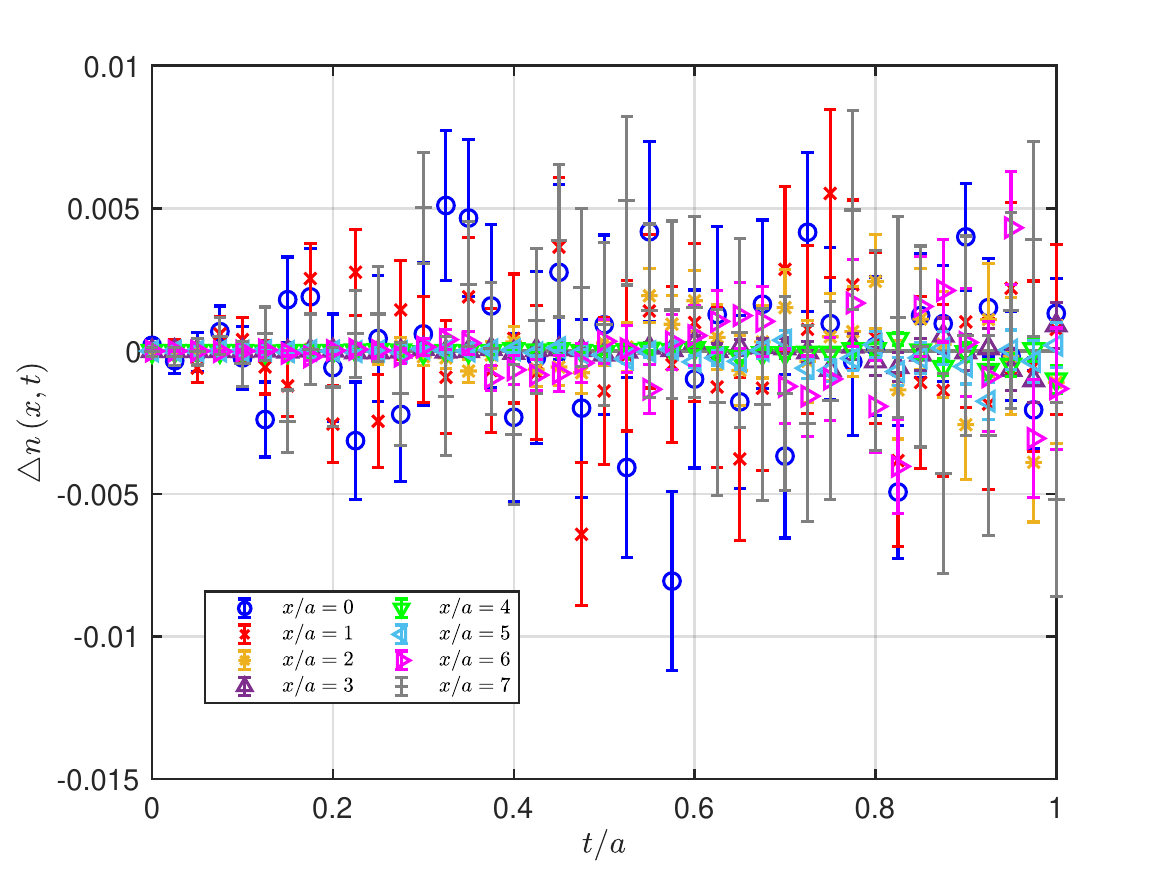}
\includegraphics[width=0.48\hsize]{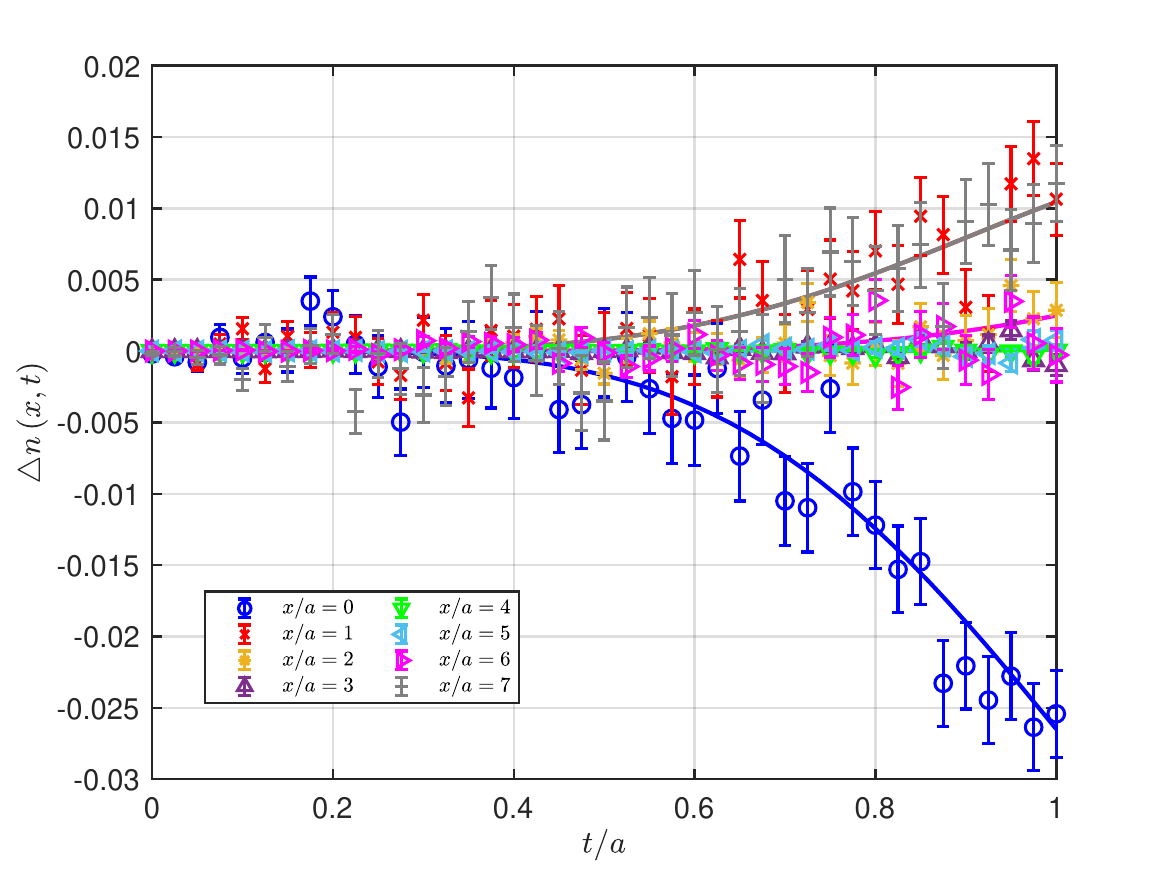}\\
\caption{\label{fig:deltafermionnumberstate1}$\Delta n(x,t)$ as functions of time evolution, for the initial state $|1\rangle$.
The left panel corresponds to $m=0$, and the right corresponds to $am=1$, respectively.}
\end{figure}
\begin{figure}[htbp]
\includegraphics[width=0.48\hsize]{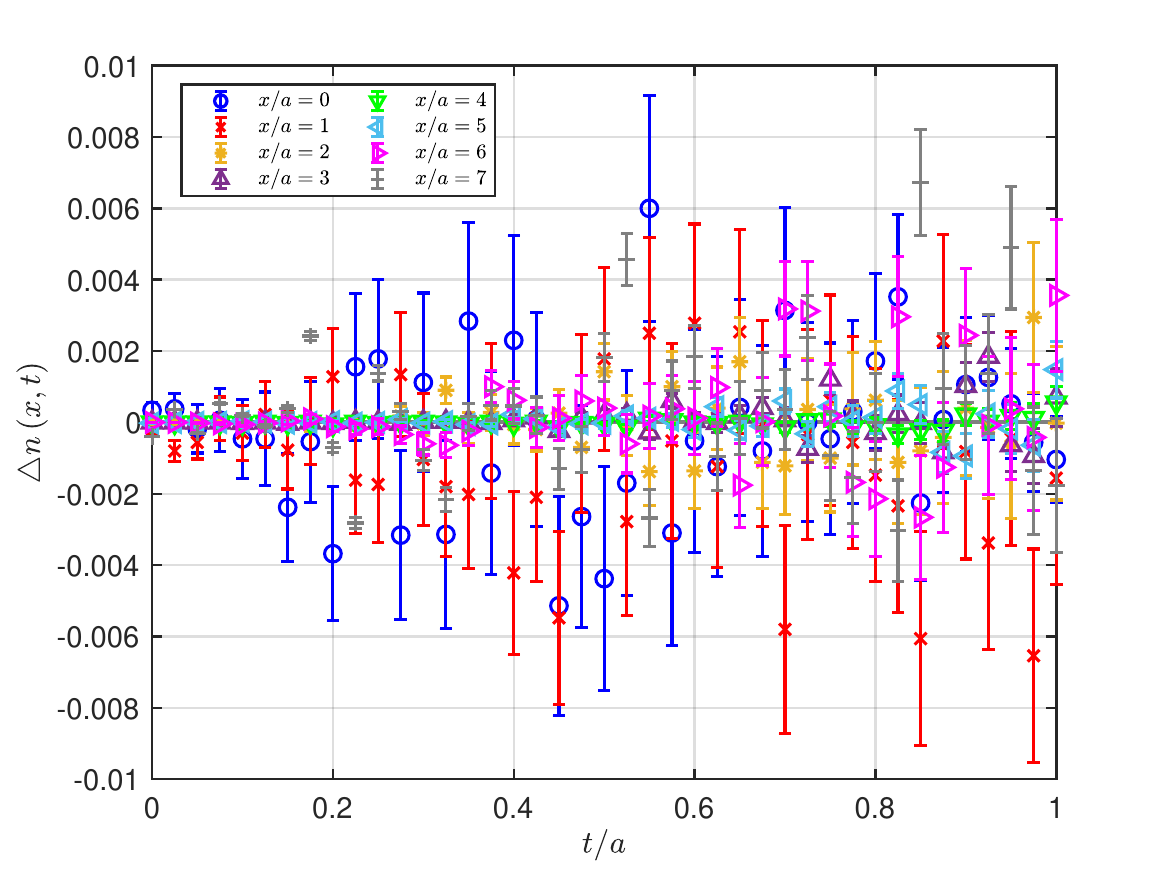}
\includegraphics[width=0.48\hsize]{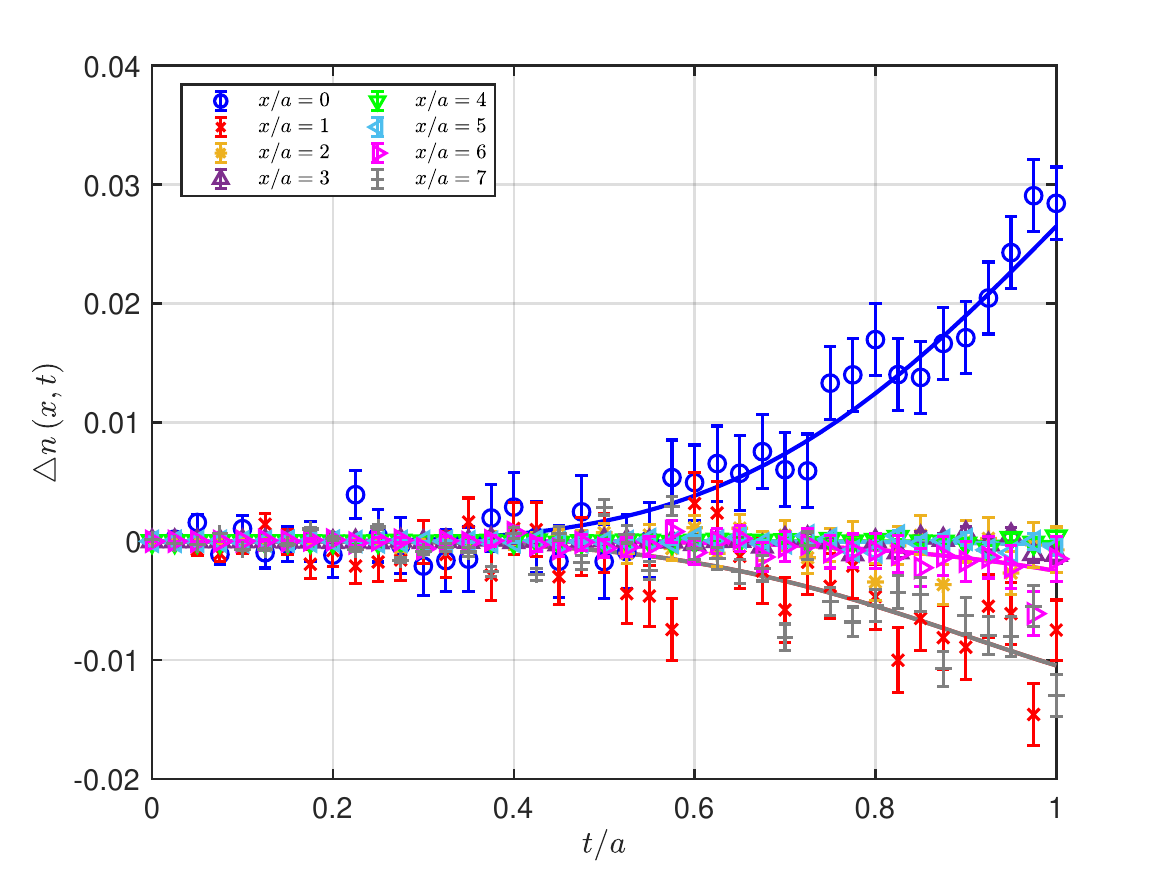}\\
\caption{\label{fig:deltafermionnumberstate2}Same as Fig.~\ref{fig:deltafermionnumberstate1} but for initial state $|254\rangle$.}
\end{figure}
Denoting $\Delta n(x,t)=n(x,t,ah=0.1)-n(x,t,ah=0)$, which reflects the effect of expanding universe, the results with initial states $|1\rangle$ and $|254\rangle$ are shown in Figs.~\ref{fig:deltafermionnumberstate1} and \ref{fig:deltafermionnumberstate2}.
It can be shown that, for massless case, the expanding universe has no effect on fermion number distribution $n(x)$.
For massless fermions in an FLRW universe, the Dirac equation is conformally invariant.
This can be shown by noting that, when $\psi(x,t)$ is a solution of the Dirac equation, then $g^{-1/2}(t)\psi(x,t)$ is a solution of Dirac equation in a flat spacetime, as shown in Appendix~\ref{sec:ap1}.

Compared with $n(x,t,h)$, the magnitude of $\Delta n(x,t)$ is about two order of magnitude smaller. 
Although wave packets spread in flat space due to dispersion, the $\Delta n(x,t)$ shows a weakened spread out, which can be recognized as a consequence of momentum redshift.
As can be seen from Fig.~\ref{fig:fermionnumberstate1}, for $|1\rangle$, the fermions at sites other than $x=0$ spread to the site $x=0$, however, Fig.~\ref{fig:deltafermionnumberstate1} shows that the growing of the fermion number at site $x=0$ is slower than the case of flat space when $m\neq 0$, which indicates a momentum redshift.
Similar phenomenon can be shown in Fig.~\ref{fig:deltafermionnumberstate2} that, the reduction of the fermion number at the site $x=0$ is smaller when $ah=0.1$ and $am=1$.

\begin{figure}[htbp]
\includegraphics[width=0.7\hsize]{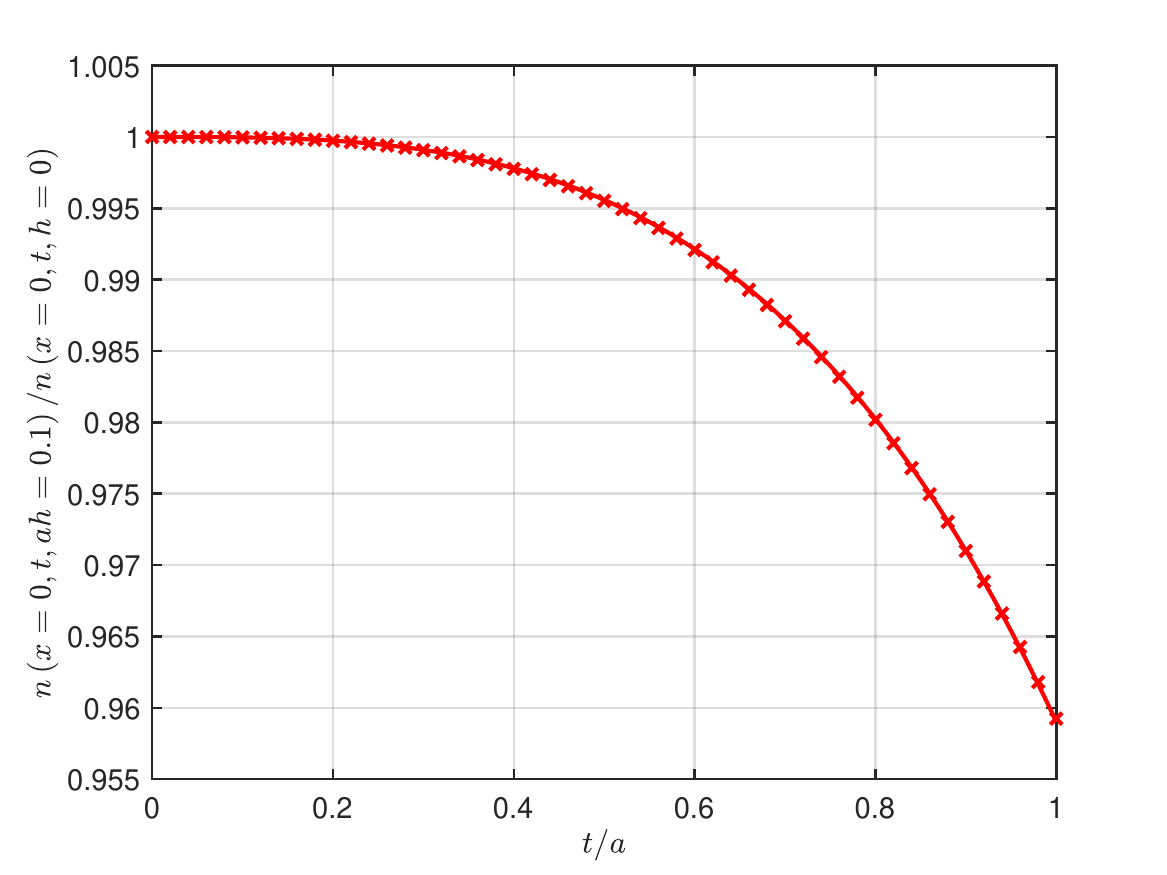}
\caption{\label{fig:fermionnumberfit}The fitting of the result of $n(x=0,t,ah=0.1)/n(x=0,t,h=0)$~(the `$\times$' points) obtained by using exact diagonalization according to $1+\alpha \left(te^{ht}\right)^3$~(the solid line).}
\end{figure}
Another observation is that, the case of $|1\rangle$ is just opposite to the case of $|254\rangle$.
This can be shown by comparing the right panels of Figs.~\ref{fig:deltafermionnumberstate1} and \ref{fig:deltafermionnumberstate2}.
The result can be understood by noticing that the charge conjugation of the $\bar{\psi}\gamma _0\psi$ term is $-1$.
Due to this symmetry, hereafter we will consider only the $|1\rangle$ state as the initial state.
To compare the result against the redshift rate $e^{ht}$, $n(x=0,t,ah=0.1)/n(x=0,t,h=0)$ is calculated by using exact diagonalization, and it is found that the result can be fitted as,
\begin{equation}
\begin{split}
&\frac{n(x=0,t,ah=0.1)}{n(x=0,t,h=0)}\approx 1+\alpha \left(te^{ht}\right)^3,
\end{split}
\label{eq.3.9}
\end{equation}
with $\alpha \approx -0.0303553$.
The fitting is shown in Fig.~\ref{fig:fermionnumberfit}.

\begin{figure}[htbp]
\includegraphics[width=0.7\hsize]{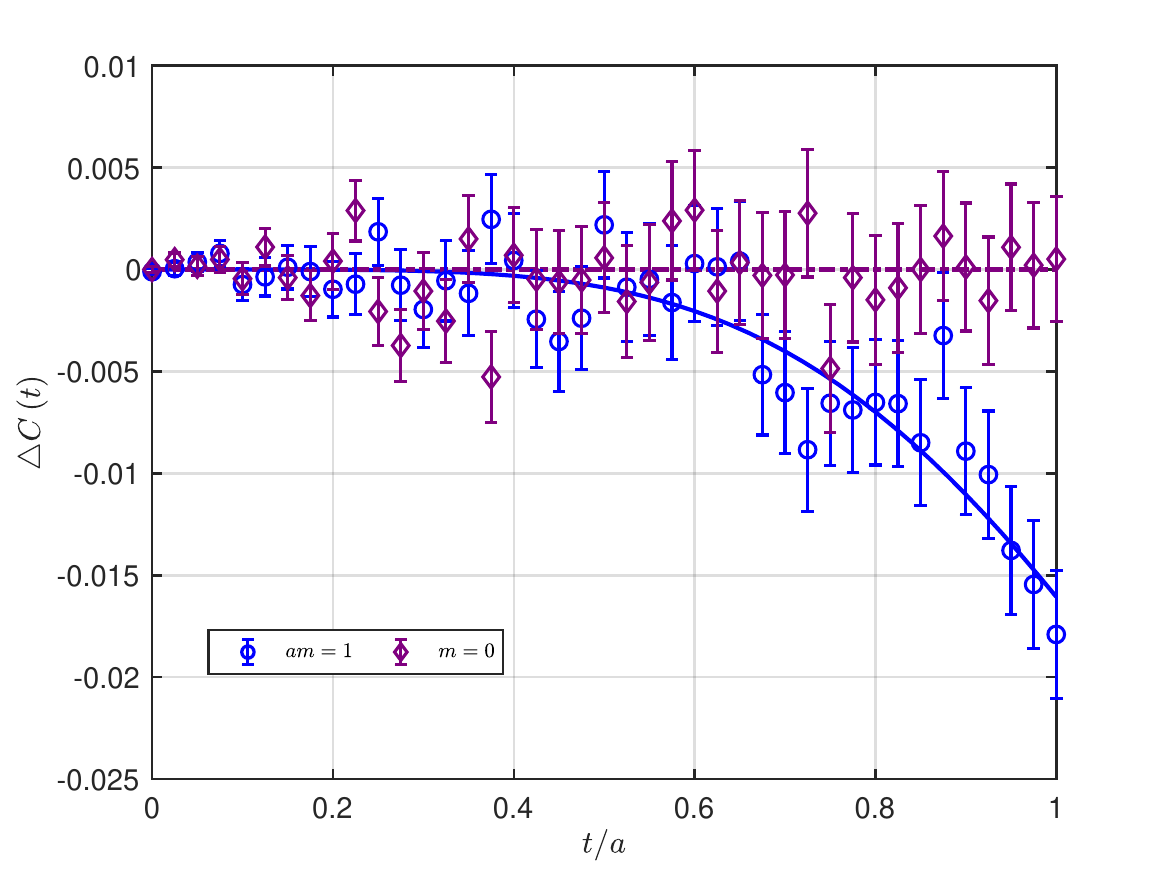}
\caption{\label{fig:deltact}$\Delta C(t)$ as function of $t$.
The dashed line and diamonds correspond to $m=0$, and the solid line and the circles correspond to $am=1$, respectively.}
\end{figure}

The density correlation between the first and second site is also studied.
Denoting $\Delta C(t)=\langle \hat{C}\rangle _{ah=0.1,t}-\langle \hat{C}\rangle _{h=0,t}$, with $\hat{C}$ defined in Eq.~(\ref{eq.3.3}), the $\Delta C(t)$ is shown in Fig.~\ref{fig:deltact}.
Again, for the massless case, no effect from the expanding universe can be found.
However, for the massive case, a decrease of density correlation is observed.
Due to fermions 'spread out' from sites where $x\neq 0$ toward $x=0$, a density correlation emerges between the sites $x=0$ and $x=1$. 
However, because redshift causes the 'spread out' process to be slower when spatial expansion presents, the correlation exhibits a slower growth compared to the case of flat space.

\begin{figure}[htbp]
\includegraphics[width=0.7\hsize]{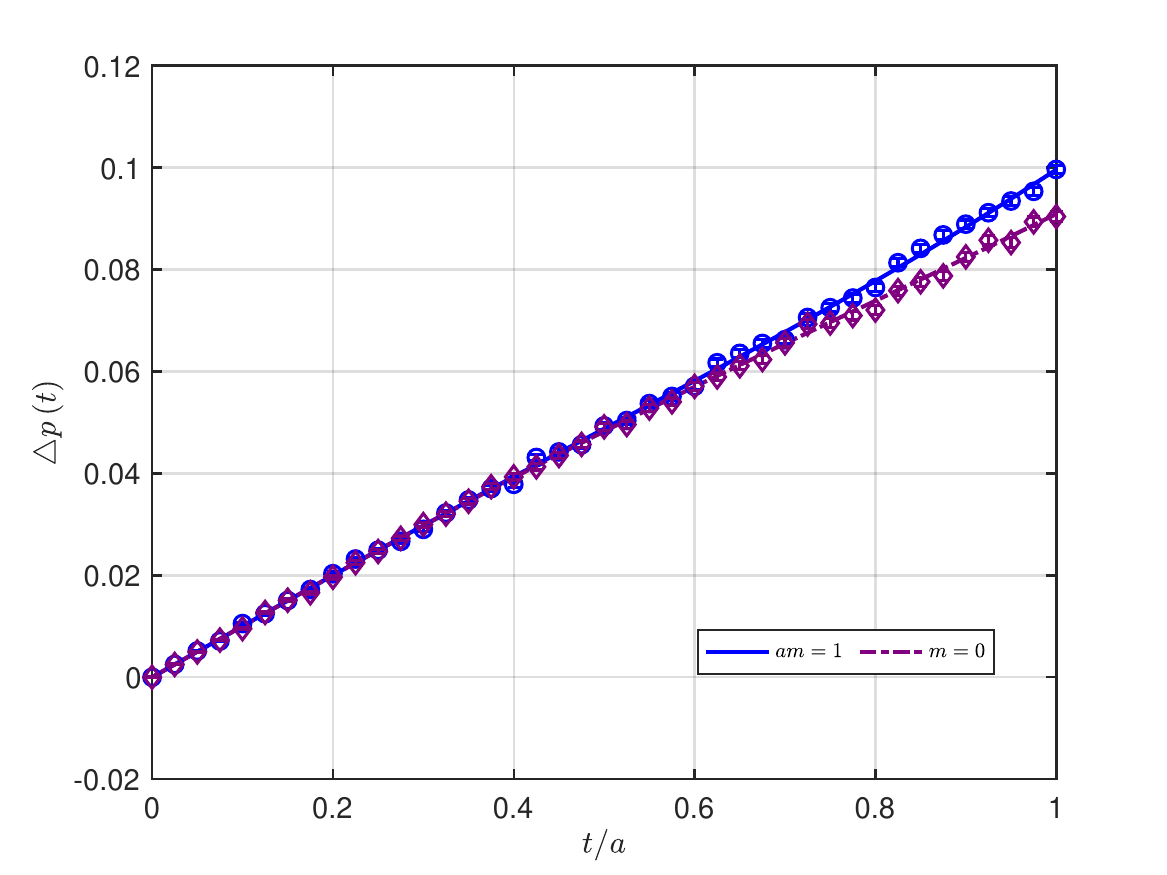}
\caption{\label{fig:polarization}Same as Fig.~\ref{fig:deltact} but for $\Delta p(t)$.}
\end{figure}
\begin{figure}[htbp]
\includegraphics[width=0.48\hsize]{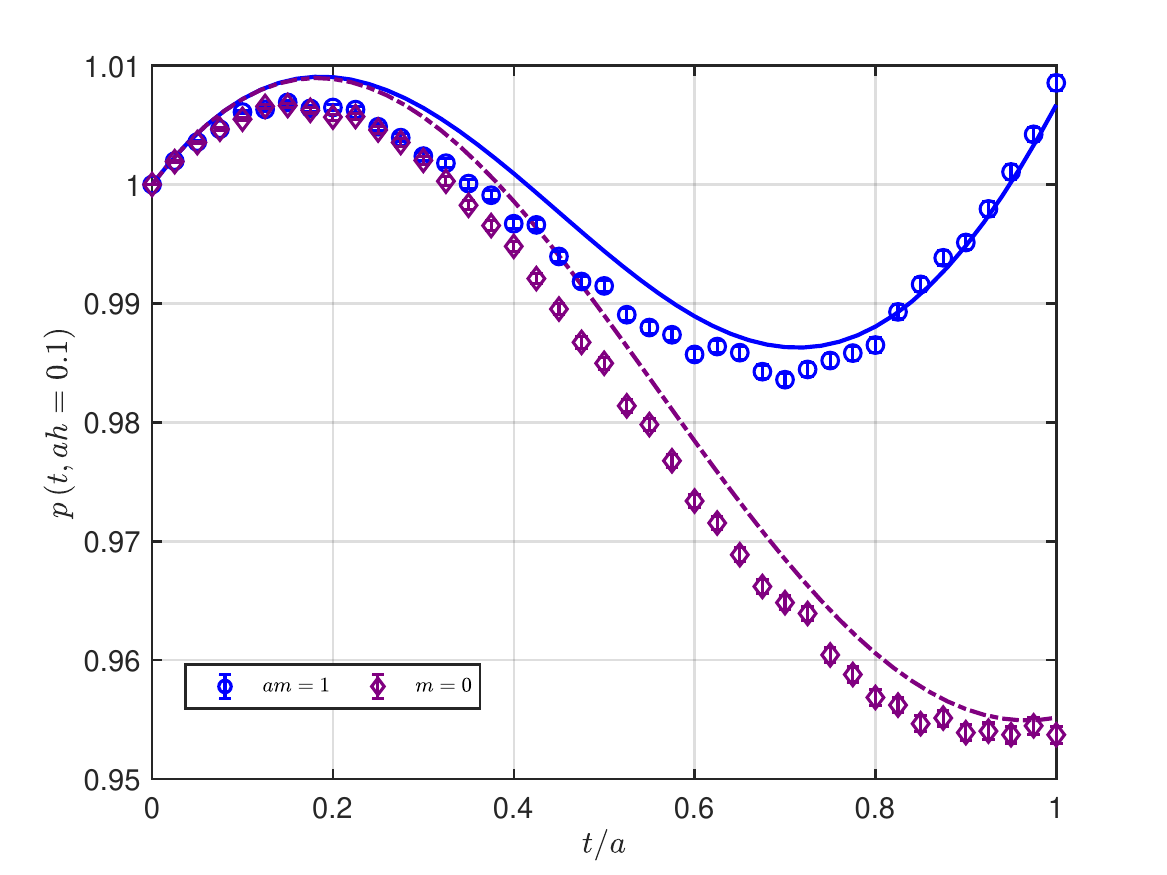}
\includegraphics[width=0.48\hsize]{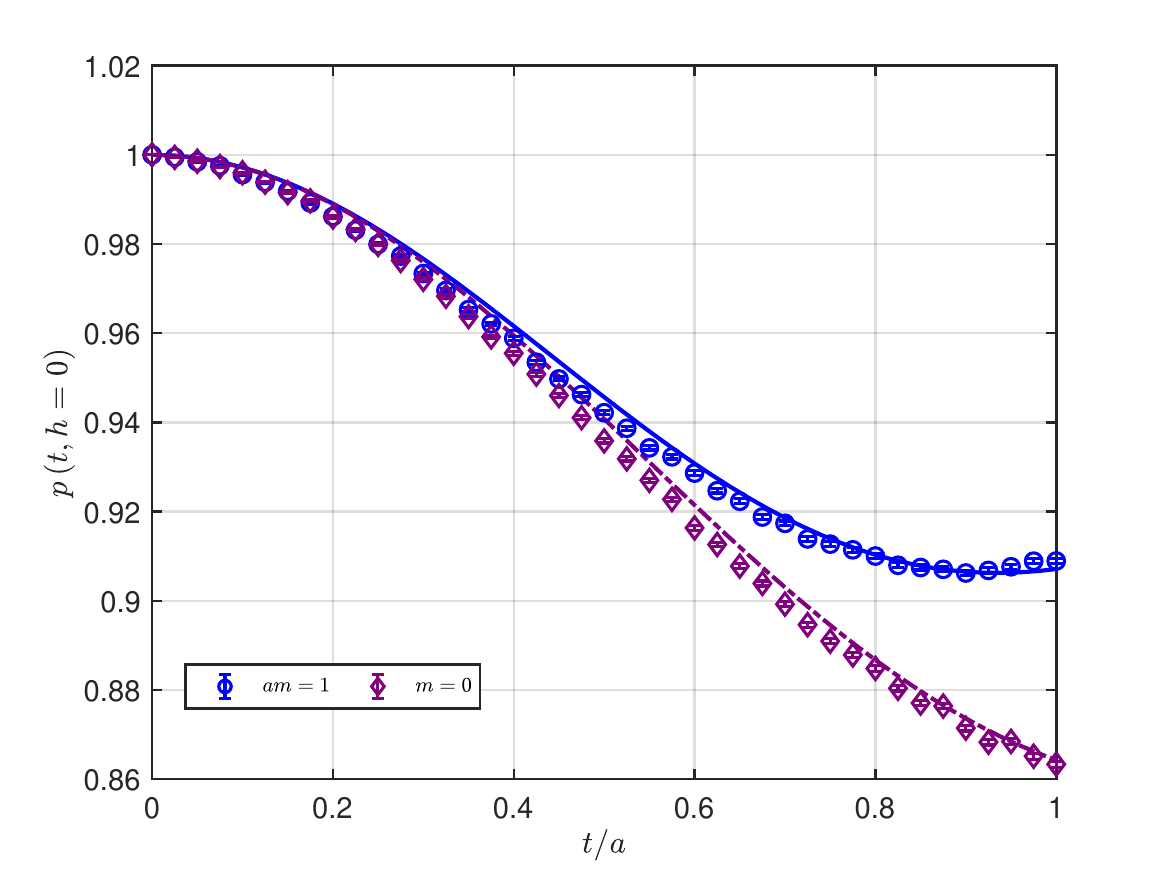}\\
\caption{\label{fig:polarization2}$p(t,ah=0.1)$~(the left panel) and $p(t,h=0)$~(the right panel) for the cases of $m=0$ and $am=0.1$.
Note that, the variation is small in the case of $ah=0.1$, the deviation is mainly due to the trotter error.}
\end{figure}

The electric polarization defined in Eq.~(\ref{eq.3.4}) is also measured.
Defining $\Delta p(t)=p(t,ah=0.1)-p(t,h=0)$, with $p(t,h)=\langle \hat{p}(t)\rangle _{t,h}/\langle \hat{p}(t)\rangle _{t=0,h} $, $\Delta p(t)$ is shown in Fig.~\ref{fig:polarization}.
An increase in electric dipole moment is observed, indicating that the redshift-induced wavefunction spreading is translated into a greater spatial extent of the charge distribution.
In the absence of cosmic expansion, the spreading of the wave packet causes fermions to diffuse from non-zero lattice sites towards the zero site, resulting in a reduction of electric polarization, as shown in the right panel of Fig.~\ref{fig:polarization2}. 
In the presence of cosmic expansion, both wave packet spreading~(which decreases polarization) and cosmic expansion~(which increases polarization) occur simultaneously. 
This leads to a competition between the two effects as shown in the left panel of Fig.~\ref{fig:polarization2}. 
As a result, the reduction of electric polarization is slower than the case of the flat space, causing $\Delta p(t)$ to increase over time.

\begin{figure}[htbp]
\includegraphics[width=0.7\hsize]{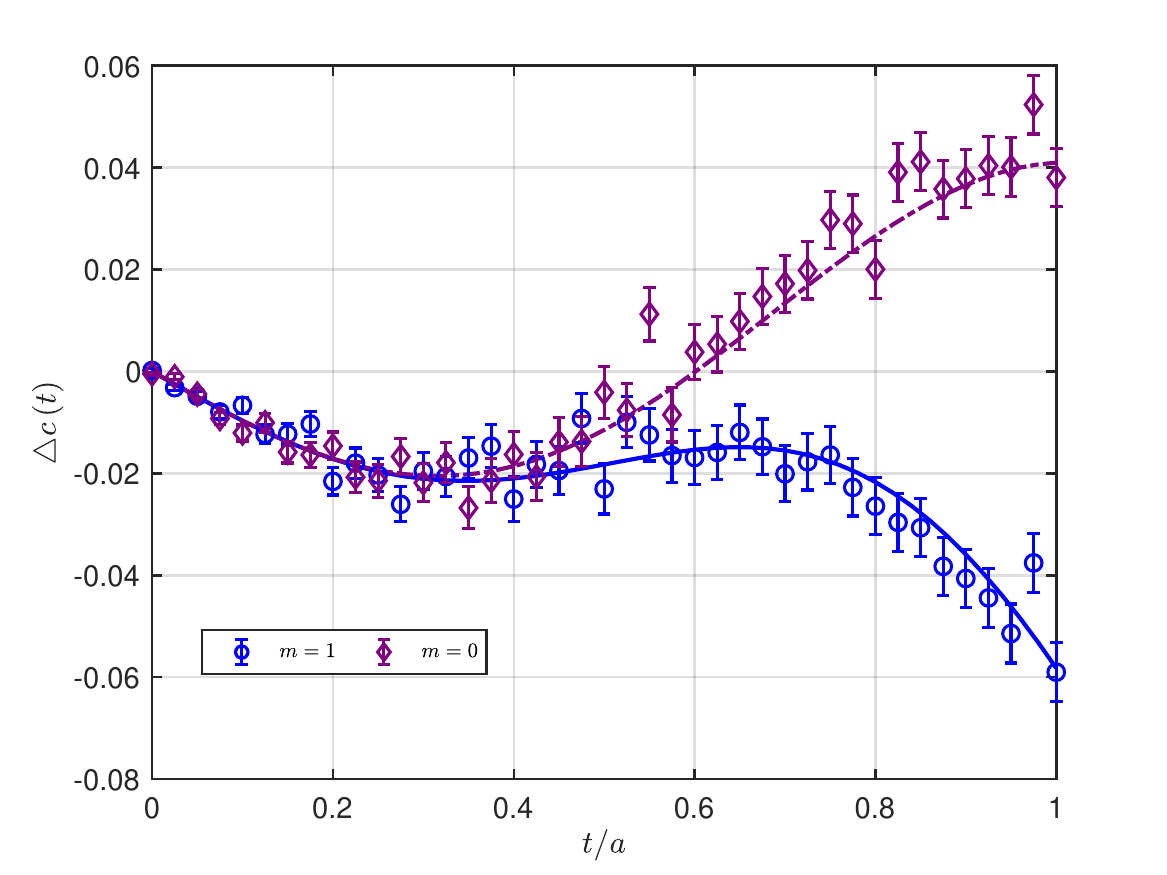}
\caption{\label{fig:deltaChiral}Same as Fig.~\ref{fig:deltact} but for $\Delta c(t)$.}
\end{figure}

Chiral condensation is a measure of the local `pairing' of left- and right-handed components. 
Denoting $\Delta c(t)=\langle c(t)\rangle _{ah=0.1,t}-\langle c(t)\rangle _{h=0,t}$, with $c(t)$ operator defined in Eq.~(\ref{eq.3.5}), $\Delta c(t)$ is shown in Fig.~\ref{fig:deltaChiral}.
For the massive case, $\Delta c(t)$ is found to decreases with time.
If the fermions become less delocalized when expanding is present due to the redshift, the strength of the above mentioned local pairing is reduced.
Not only that, for both massless case and the massive case, the chiral condensation also exhibits oscillatory behavior at larger $t$, which may be attributed to the non-equilibrium of the system.

\subsection{\label{sec:3.4}finite-volume effect}

\begin{figure}[htbp]
\includegraphics[width=0.7\hsize]{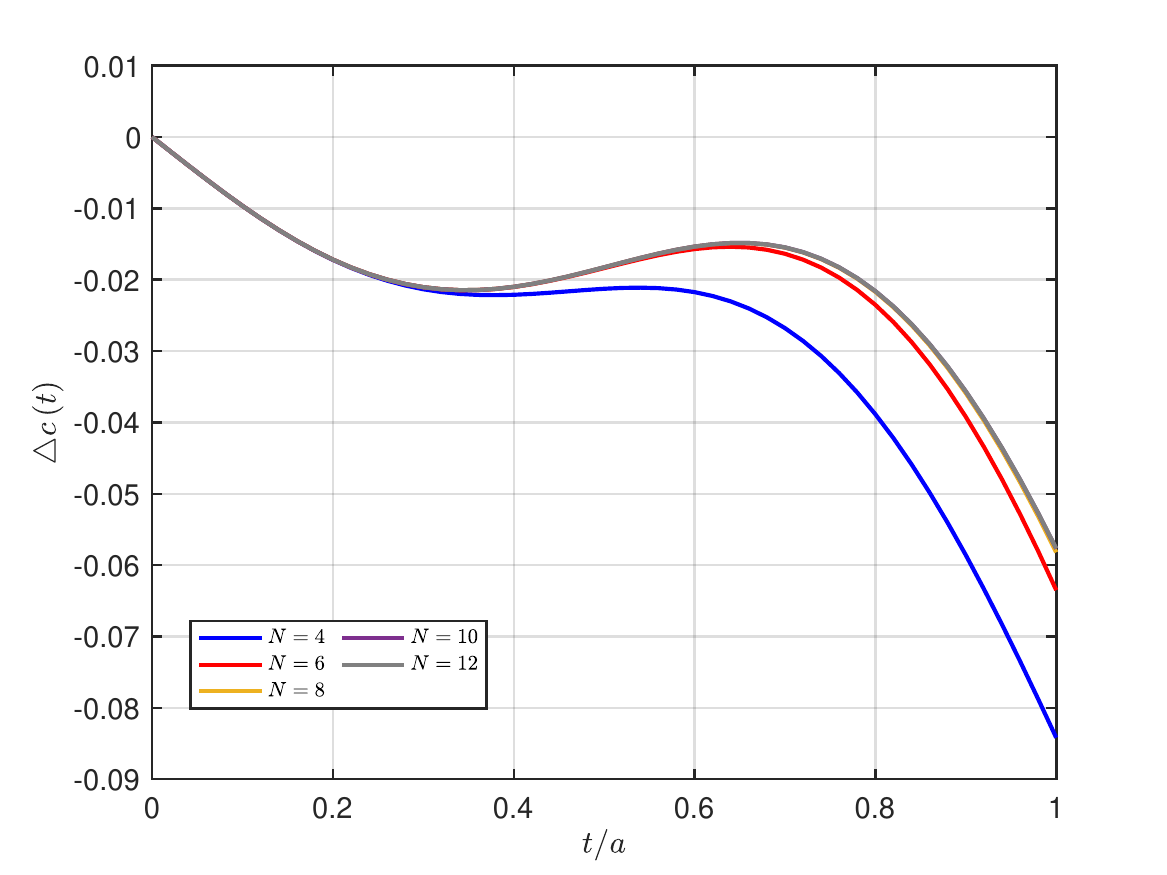}
\caption{\label{fig:finitevolume}$\Delta c(t)$ at $N=4$, $6$, $8$, $10$, and $12$ calculated by using exact diagonalization.}
\end{figure}

\begin{figure}[htbp]
\includegraphics[width=0.7\hsize]{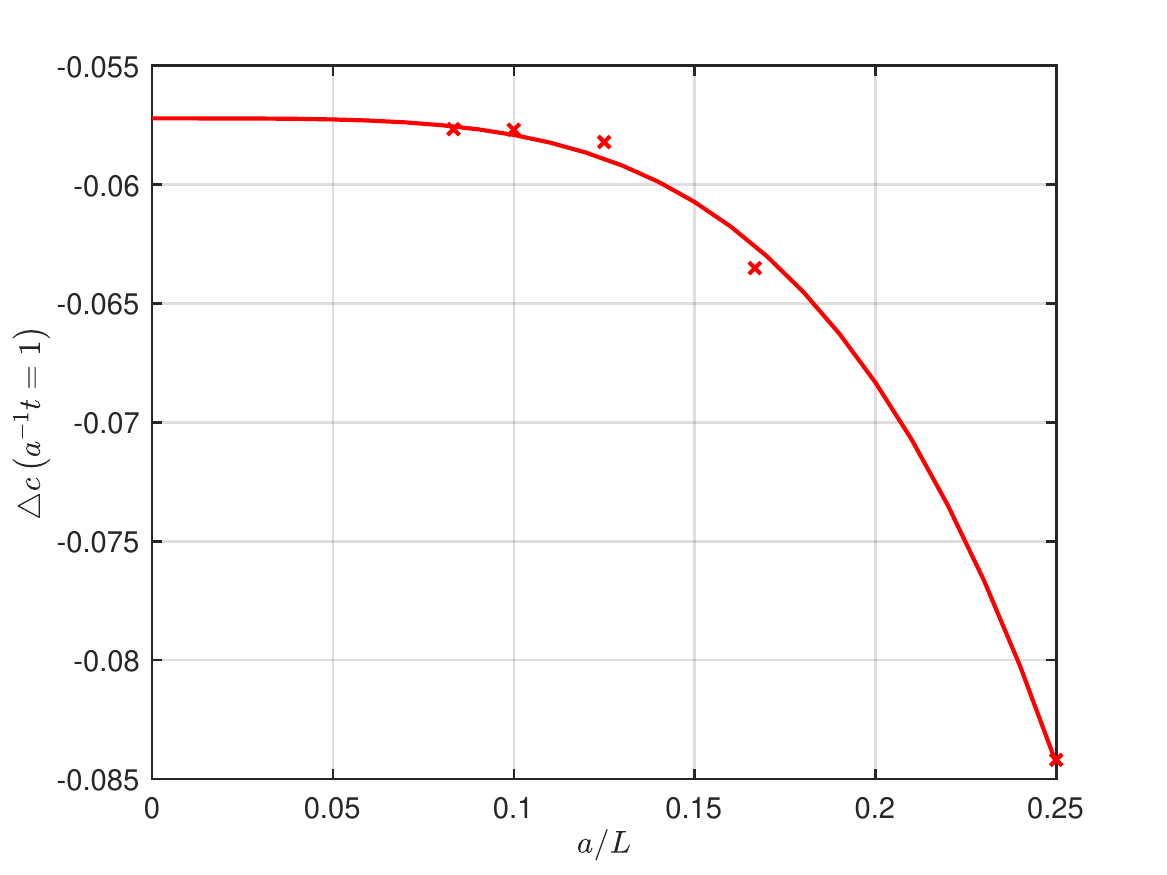}
\caption{\label{fig:finitevolume2}$\Delta c(t=1a)$ at $N=4$, $6$, $8$, $10$, and $12$ calculated by using exact diagonalization. The solid line is a fitting according to $\alpha + \beta (a/L)^4$ which serves as a guide to the eye.}
\end{figure}

Since $N=8$ is relatively small, our system may be affected by finite-volume effects. 
Therefore, the numerical results at different $N$ are studied. 
Focusing on the case of $am=1$ and the chiral condensation, for the initial state $|1\rangle$, by using exact diagonalization, the results for $\Delta c(t)$ at $N=4$, $6$, $8$, $10$, and $12$ are calculated.
The results are shown in Fig.~\ref{fig:finitevolume}.
It can be seen that, the cases of $N=8$, $10$, and $12$ are close to each other.
Assuming that the contribution from finite-volume effects is a function of $a/L$~(where $L=aN$ is the system size), the dependence of $\Delta c(t)$ on $a/L$ at $a^{-1}t=1$ is shown in Fig.~\ref{fig:finitevolume2}~(the solid line serves as a guide to the eye). 
It can be observed that $\Delta c (t)$ varies nonlinearly as $a/L$ decreases. 
The finite-volume effect (denoted as $fv$) can be estimated as,
\begin{equation}
\begin{split}
&fv<\frac{\frac{1}{8}-\frac{1}{12}}{\frac{1}{8}}\left|\Delta c (t)_{N=12}-\Delta c (t)_{N=8}\right|.
\end{split}
\label{eq.3.10}
\end{equation}
At $a^{-1}t=1$, this effect amounts to approximately $0.3\%$, therefore we can conclude that the finite-volume effect is small.

Also note that, without periodic boundary condition, the Hamiltonian is two-local, and two-qubit gates would solely originate from the kinetic term. 
When arranging these gates in a pairwise manner while ignoring single-qubit gates, each timestep $\Delta t$ requires $4$ circuit layers. 
With $K=40$, the number of layers is $160$, and increasing system size on actual quantum devices will not increase circuit depth. 
Unlike the case of classical simulation, when quantum error correction or hardware advancements enable maintaining fidelity for such a quantum circuit, increasing system size on actual quantum devices will not increase time computational complexity, and will only in increase the number of gate as a logarithm function of the system size. 
As a result, simulating large-scale systems is precisely where the advantage of quantum simulation lies.

\subsection{\label{sec:3.5}Systematic errors}

In our simulations, errors can be categorized into the following components, (i) error arising from the finite number of measurements, classified as statistical error; (ii) error induced by noise; (iii) error caused by temporal discretization; (iv) error originating from Trotter decomposition. 
The error due to finite measurements is given in Eq.~(\ref{eq.3.7}).

Our simulations employ periodic boundary conditions. 
As discussed above, without periodic boundaries, each timestep $\Delta t$ requires $4$ circuit layers. 
Given current two-qubit quantum gate error rates of $2-5\%$, results become unreliable after approximately $5$ timesteps. 
The difference between simulations with and without cosmic expansion is small, which requires larger timesteps. 
We therefore conclude that implementation on current-stage quantum hardware remains impractical without error correction, and omit further noise analysis.

\begin{figure}[htbp]
\includegraphics[width=0.48\hsize]{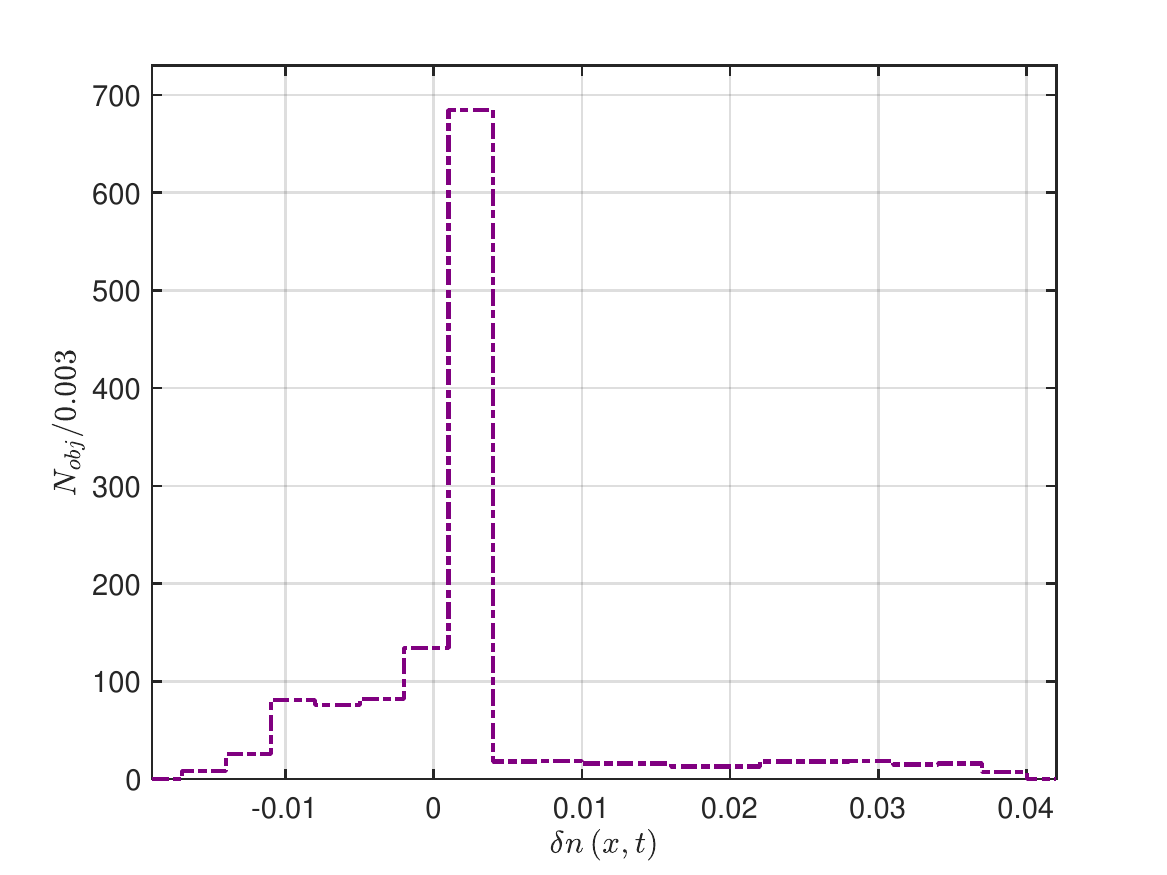}
\includegraphics[width=0.48\hsize]{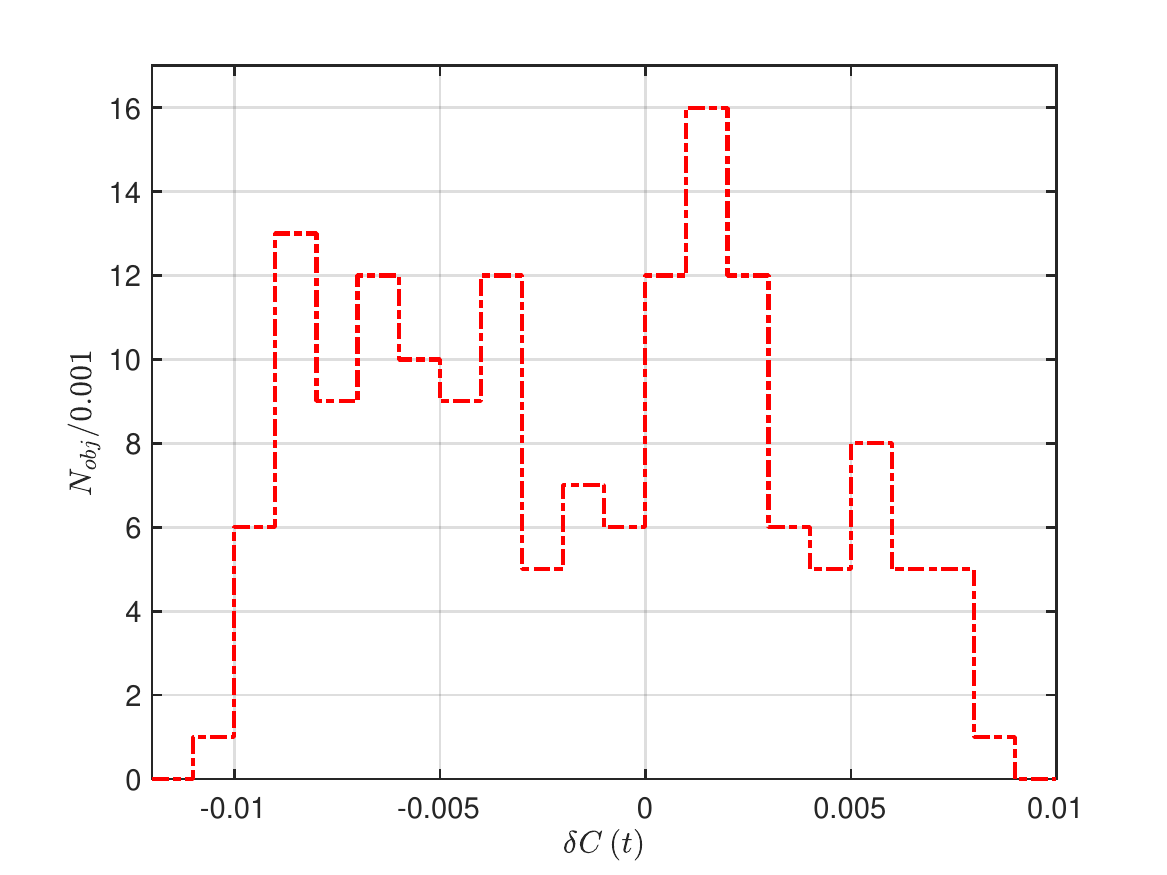}\\
\includegraphics[width=0.48\hsize]{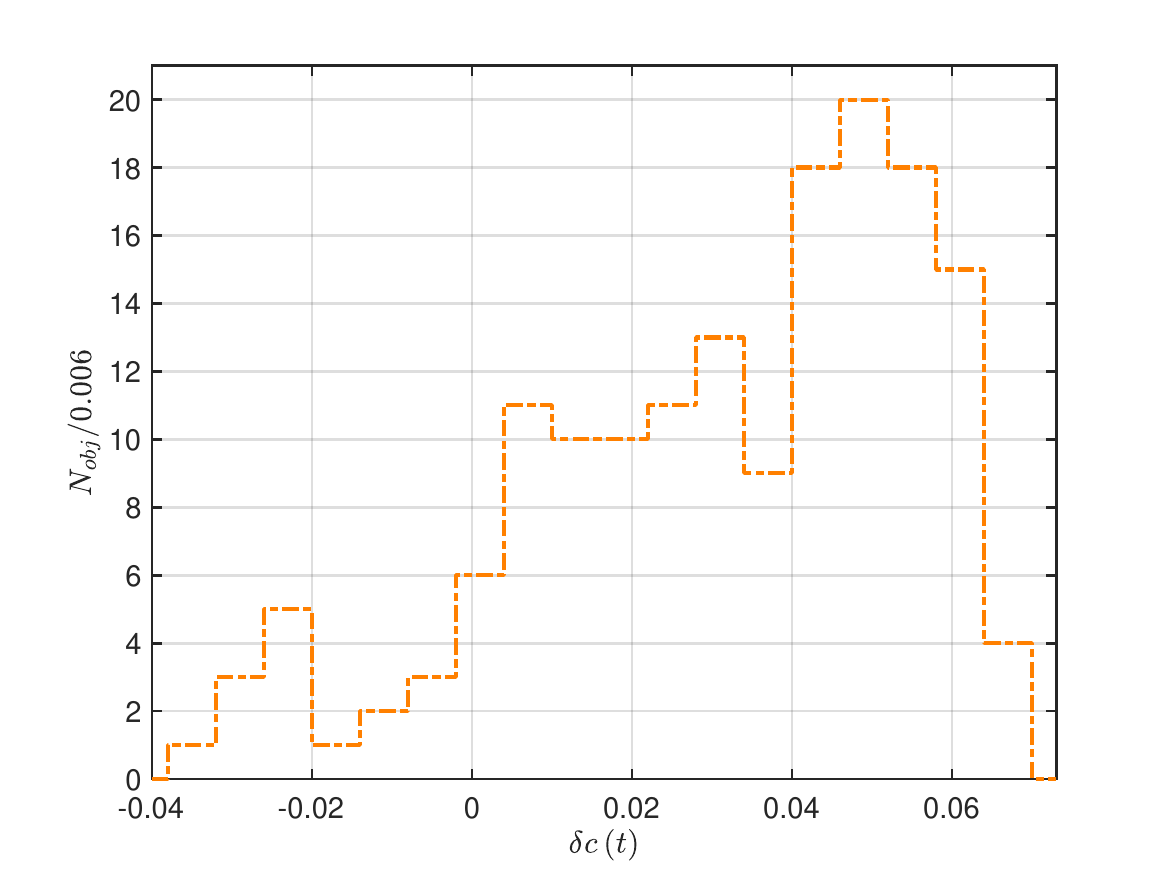}
\includegraphics[width=0.48\hsize]{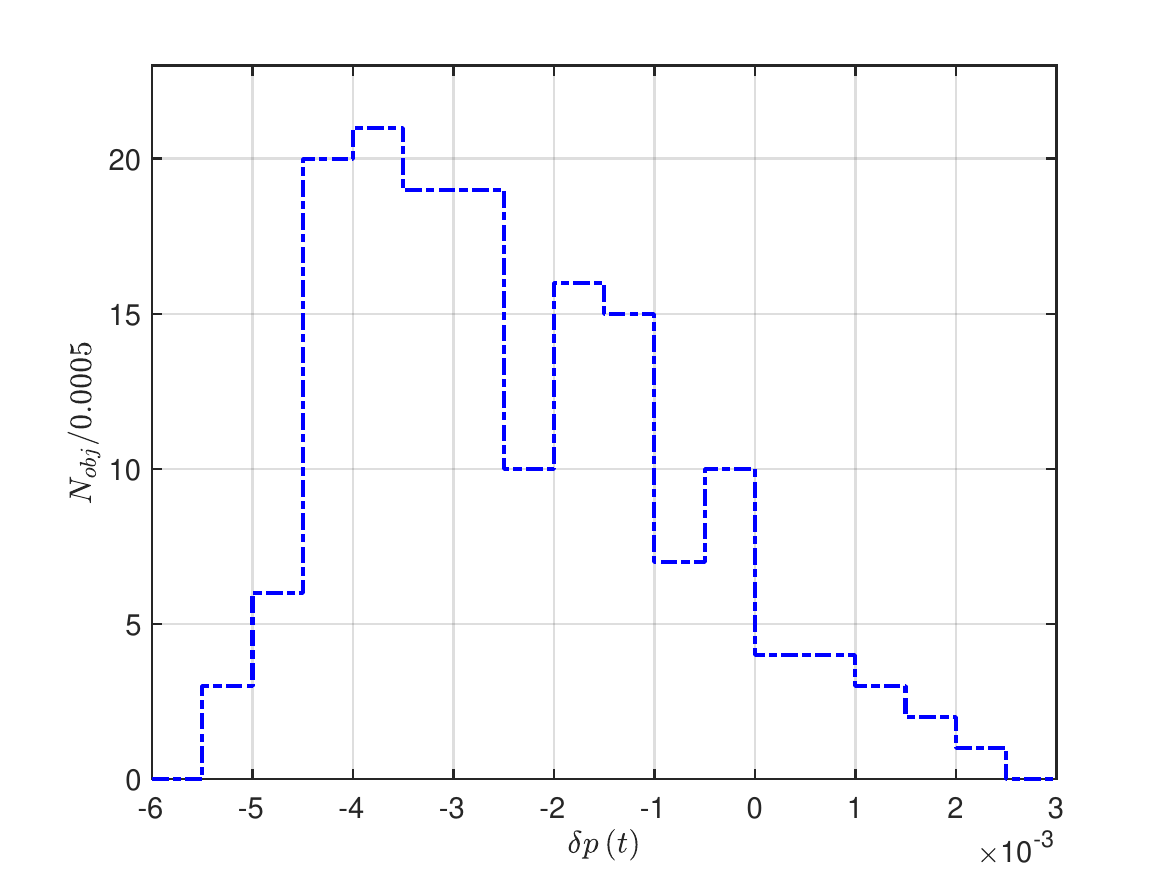}\\
\caption{\label{fig:dobserv}The distributions of $\delta n$, $\delta C$, $\delta p$, and $\delta c$.}
\end{figure}
Except for noise, errors from (iii) and (iv) are both at the order of $\mathcal{O}(\Delta t)$. 
One can quantify the magnitude of errors from (i), (iii) and (iv) by comparing results from exact diagonalization with those obtained from quantum circuits.
Denoting $\delta O$ as $\langle \hat{O} \rangle _{t,h}-\langle \hat{O}\rangle _{ex}$, $\langle \hat{O}\rangle _{ex}$ means the result obtained by using exact diagonalization, the distributions of $\delta n$, $\delta C$, $\delta p$, and $\delta c$~(in the definition of $\delta p$, we use normalized $\langle \hat{p}(t) \rangle _{t,h}/\langle \hat{p}(t) \rangle _{t=0,h}$ as in section~\ref{sec:3.3}) are shown in Fig.~\ref{fig:dobserv}.
For $\delta n$, there are $N\times K$ points for each $m$ and $h$, therefore $N_{obj}=8\times 40 \times 4 = 1280$ points, for the other cases there are $K$ points for each $m$ and $h$, therefore $N_{obj}=160$ points.
All observables are functions of $\hat{n}(x)$. 
Focusing on $\hat{n}(x)$, it can be observed that its errors cluster near zero with a slight positive bias, remaining within $0.04$.

The requirement for the fidelity threshold depends on two aspects. 
On one hand, as shown in Eq.~(\ref{eq.3.9}), for smaller values of $h$, higher precision is required to observe differences between the presence and absence of expansion. 
On the other hand, it depends on the required time $t$. 
Assuming a larger $h$, such that a significant difference between the presence and absence of cosmic expansion can be observed at $t=1$, we can use state fidelity to estimate the fidelity requirement for two-qubit gates. 
For a circuit depth of $160$~(without the periodic boundary condition and with $K=40$), if the result fidelity is required to be greater than $80\%$, then $(1 - \epsilon)^{160} > 0.8$, where $\epsilon$ is the two-qubit gate error rate, which must be less than $0.14\%$.
The current two-qubit gate error rate of Zuchongzhi 2.0 is about $1.1\%$, indicating a gap of one order of magnitude. 
However, modern error mitigation techniques may significantly relax the fidelity requirements, making it possible to obtain reliable results from a circuits with depths up to $60$~\cite{Kim:2021gvc}. 
Along with the tensor network circuit optimization technique~\cite{Chai:2025kbi}, we believe this work can be implemented on a real quantum computer in the near future.

\section{\label{sec:4}Summary}

This work explores the application of quantum computing in simulating fermion field dynamics in an expanding universe, the $1+1$ dimensional de Sitter spacetime. 
By leveraging digital quantum simulations, this work implements the Jordan-Wigner transformation and Trotter decomposition to simulate fermionic behavior on a universal quantum computer.

The simulation is carried out on a simulator by using the \verb"Qiskit" package.
The evolution of the fermion number density distribution over time is investigated and a spread out phenomenon is observed, which is a consequence of momentum redshift. 
In addition, the fermion density correlation function, and the chiral condensation are studied, and are found to be coinciding with the spread out of fermion density. 
The fermion polarization is also studied and used as a quantity to test the error.

Future studies can extend this approach by incorporating interactions, such as gauge fields, to explore quantum electrodynamics in curved spacetime. 
Additionally, studying higher-dimensional models or implementing error correction strategies could enhance the accuracy and scalability of quantum simulations. 
These advancements would provide deeper insights into early-universe phenomena and quantum gravity effects.

\begin{acknowledgements}
This work was supported in part by the National Natural Science Foundation of China under Grants No.~12147214, the Basic Research Projects of Universities in Liaoning Province (Grant Nos.~LJ212510165024 and LJKMZ20221431).
\end{acknowledgements}

\section*{Data Availability}
The data and code that support the findings of this study are available at \url{https://www.modelscope.cn/datasets/nbalexis/Collision_events_of_aavv_final_states_with_aQGCs_at_muon_colliders}~\cite{data}.

\appendix

\section{\label{sec:ap1}Fermion Hamiltonian in the FLRW space}

One can start with the Dirac equation in $1+1$ dimension curved space,
\begin{equation}
\begin{split}
&\left(i \gamma^{\mu} \partial_{\mu}+i \gamma^{\mu} \Gamma_{\mu}-m\right) \psi=0,\\
\end{split}
\label{eq.ap.1}
\end{equation}
where, the spin connection is,
\begin{equation}
\Gamma_{\mu}=\frac{1}{4} \sigma^{i j} \omega_{\mu i j},
\label{eq.ap.2}
\end{equation}
with,
\begin{equation}
\begin{split}
&\sigma^{i j}=\frac{i}{2}\left[\gamma^{i}, \gamma^{j}\right],\\
&\omega_{\mu i j}=g_{\alpha \beta} e_{i}^{\alpha}\left(\partial_{\mu} e_{j}^{\beta}+\Gamma_{\mu \nu}^{\beta} e_{j}^{\nu}\right),\\
\end{split}
\label{eq.ap.3}
\end{equation}
where $g_{\alpha\beta}$ is the metric tensor,
\begin{equation}
g_{\mu \nu}=\left(\begin{array}{cc}
1 & 0  \\
0 & -g^{2}(t)
\end{array}\right),
\label{eq.ap.4}
\end{equation}
$\Gamma_{\mu \nu}^{\beta}$ are the Christoffel symbols,
\begin{equation}
\Gamma_{\mu \nu}^{\beta}=\frac{1}{2} g^{\beta \rho}\left(\partial_{\mu} g_{\rho \nu}+\partial_{\nu} g_{\mu \rho}-\partial_{\rho} g_{\mu \nu}\right),
\label{eq.ap.5}
\end{equation}
and $e_{\mu}^i$ are vielbein,
\begin{equation}
\begin{cases}
    e_{t}=(1,0)\\
    e_{x}=\left(0, g^{-1}(t)\right)
\end{cases}, \quad \begin{cases}
    e^{t}=(1,0)\\
    e^{x}=(0, g(t))
\end{cases}.
\label{eq.ap.6}
\end{equation}

It can be verified that non-zero Christoffel symbols are,
\begin{equation}
\begin{split}
&\Gamma_{x x}^{t}=g(t) g^{\prime}(t), \\
&\Gamma_{t x}^{x}=\Gamma_{x t}^{x}=\frac{g^{\prime}(t)}{g(t)},
\label{eq.ap.7}
\end{split}
\end{equation}
which result in non-zero $\omega$,
\begin{equation}
\begin{split}
& \omega_{x, x t}=-g^{\prime}(t),\;\;\;
  \omega_{x, t x}=g^{\prime}(t),
\end{split}
\label{eq.ap.8}
\end{equation}
and therefore the non-zero spin connection term,
\begin{equation}
\begin{split}
&\Gamma_{x}=-\frac{1}{2} g^{\prime}(t) \sigma^{x t}.
\end{split}
\label{eq.ap.9}
\end{equation}
The gamma matrices in $\sigma^{xt}$ are,
\begin{equation}
\gamma^{\mu}=\gamma^{i} e_{i}^{\mu} 
\label{eq.ap.10}
\end{equation}

Multiply a $\gamma ^t$ on the left of Eq.~(\ref{eq.ap.1}), the Dirac equation is then,
\begin{equation}
\left(i \partial_{t}+i \gamma^{t} \gamma^{x} \partial_{x}+i \gamma^{t} \gamma^{x} \Gamma_{x}-\gamma^{t} m\right) \psi=0,
\label{eq.ap.11}
\end{equation}
which is the Schrodinger equation $i\partial _t \psi = \hat{H}\psi$, and the Hamiltonian operator can be read out as,
\begin{equation}
\begin{split}
&\hat{H}=\left(-i \gamma^{t} \gamma^{x} \partial_{x}-i \gamma^{t} \gamma^{x} \Gamma_{x}+\gamma^{t} m\right).\\
\end{split}
\label{eq.ap.12}
\end{equation}
Then the Hamiltonian is,
\begin{equation}
\begin{split}
H&=\int d x \sqrt{-\operatorname{det}(g_{\mu\nu})}\\
&\times \psi^{\dagger}\left(-i \gamma^{t} \gamma^{x} \partial_{x}-i \gamma^{t} \gamma^{x} \Gamma_{x}+\gamma^{t} m\right) \psi\\
&=\int d x \bar{\psi}\left(-i \gamma^{1} \partial_{x}+\frac{1}{2}\frac{g'(t)}{g(t)} \gamma^{0}+g(t) m\right) \psi,
\end{split}
\label{eq.ap.13}
\end{equation}
which can be discretized as,
\begin{equation}
H=a \sum_{x} \bar{\psi}\left(-i \gamma^{1} \partial_{x}+\frac{1}{2}\frac{g'(t)}{g(t)}\gamma^{0}+g(t) m\right) \psi, 
\label{eq.ap.14}
\end{equation}
where $a$ is the lattice spacing.

Note that, in the massless case, Eq.~(\ref{eq.ap.11}) is,
\begin{equation}
\left(i \gamma ^0 \partial_{t} +i \gamma^{1} \partial_{x}-\frac{1}{2}\frac{g'(t)}{g(t)}\gamma^{0}\right) \psi=0.
\label{eq.ap.15}
\end{equation}
So, if $\psi(x,t)$ is a solution of Eq.~(\ref{eq.ap.15}), let $\phi(x,t)=\psi (x,t)/\sqrt{g(t)}$, then,
\begin{equation}
\left(i \gamma ^0 \partial_{t} +i \gamma^{1} \partial_{x}\right) \phi=0,
\label{eq.ap.16}
\end{equation}
such that $\phi$ is a solution of massless Dirac equation in a flat spacetime.

\section{\label{sec:ap2}Detail form of the Hamiltonian when \texorpdfstring{$N=8$}{N=8}}

According to Eq.~(\ref{eq.2.8}), when $N=8$, the Hamiltonian can be written as, 
\begin{equation}
\begin{split}
&aH = -h_{1}+\left ( \frac{1}{2}h\right )\times h_{2} +\left ( e^{ht}m \right )\times h_{3}
\end{split}
\label{eq.ap.17}
\end{equation}
with,
\begin{equation}
\begin{split}
&h_{1}=\frac{1}{2}\left \{\sigma ^{x}\left ( 7 \right )\sigma ^{x}\left ( 6 \right )  + \sigma ^{x}\left ( 6 \right )\sigma ^{x}\left ( 5 \right )  + \sigma ^{x}\left ( 5 \right )\sigma ^{x}\left ( 4 \right )  \right.\\
&\left.+ \sigma ^{x}\left ( 4 \right )\sigma ^{x}\left ( 3 \right )  + \sigma ^{x}\left ( 3 \right )\sigma ^{x}\left ( 2 \right )  +\sigma ^{x}\left ( 2 \right )\sigma ^{x}\left ( 1 \right )\right.\\
&\left.+\sigma ^{x}\left ( 1 \right )\sigma ^{x}\left ( 0 \right ) \right.\\
&\left.+\sigma ^{y}\left ( 7 \right )\sigma ^{y}\left ( 6 \right )  + \sigma ^{y}\left ( 6 \right )\sigma ^{y}\left ( 5 \right )  + \sigma ^{y}\left ( 5 \right )\sigma ^{y}\left ( 4 \right )  \right.\\
&\left.+ \sigma ^{y}\left ( 4 \right )\sigma ^{y}\left ( 3 \right )   + \sigma ^{y}\left ( 3 \right )\sigma ^{y}\left ( 2 \right )  +\sigma ^{y}\left ( 2 \right )\sigma ^{y}\left ( 1 \right )\right.\\
&\left.+\sigma ^{y}\left ( 1 \right )\sigma ^{y}\left ( 0 \right ) \right.\\
&\left.+\sigma ^{x}\left ( 7 \right )\sigma ^{z}\left ( 6 \right ) \sigma ^{z}\left ( 5\right )\sigma ^{z}\left ( 4 \right )\sigma ^{z}\left ( 3 \right )\sigma ^{z}\left ( 2 \right ) \sigma ^{z}\left ( 1 \right )\sigma ^{x}\left ( 0\right ) \right.\\
&\left. + \sigma ^{y}\left ( 7 \right )\sigma ^{z}\left ( 6 \right ) \sigma ^{z}\left ( 5\right )\sigma ^{z}\left ( 4 \right )\sigma ^{z}\left ( 3 \right )\sigma ^{z}\left ( 2 \right ) \sigma ^{z}\left ( 1 \right )\sigma ^{y}\left ( 0\right ) \right \},\\
\end{split}
\label{eq.ap.18}
\end{equation}
\begin{equation}
\begin{split}
&h_{2}=\frac{1}{2}\left \{ \sigma ^{z} \left ( 7 \right )+\sigma ^{z} \left ( 6 \right )+ \sigma ^{z} \left ( 5 \right ) + \sigma ^{z} \left ( 4 \right )+\sigma ^{z} \left ( 3 \right )\right.\\
&\left. +\sigma ^{z} \left ( 2 \right )+ \sigma ^{z} \left ( 1 \right ) + \sigma ^{z} \left ( 0 \right )\right \},\\
\end{split}
\label{eq.ap.19}
\end{equation}
and
\begin{equation}
\begin{split}
&h_{3}=\frac{1}{2}\left \{ -\sigma ^{z} \left ( 7 \right )+\sigma ^{z} \left ( 6 \right )- \sigma ^{z} \left ( 5 \right ) + \sigma ^{z} \left ( 4 \right )-\sigma ^{z} \left ( 3 \right )\right.\\
&\left. +\sigma ^{z} \left ( 2 \right )- \sigma ^{z} \left ( 1 \right ) + \sigma ^{z} \left ( 0 \right )\right \}.\\
\end{split}
\label{eq.ap.20}
\end{equation}

\smallskip

\smallskip

\bibliography{de-sitter}

\end{document}